\documentclass[useAMS,usenatbib]{mn2e}

\usepackage{natbib}
\usepackage{graphicx}
\usepackage{color}
\usepackage{amssymb}
\usepackage{amsmath}
\usepackage{xspace}
\usepackage{xargs}
\usepackage{url}

\usepackage[usenames,dvipsnames]{xcolor}
\definecolor{FireBrick}{RGB}{178, 34, 34}

\usepackage[usenames,dvipsnames]{xcolor}

\newcommandx{\N}[2][1= ,2= ]{$\mathcal{N}^{#1}_{#2}$\xspace}
\newcommandx{\R}[2][1= ,2= ]{$\mathcal{R}^{#1}_{#2}$\xspace}

\begin{document}
\title[Galaxy Zoo: spiral arm number and debiasing]{Galaxy Zoo: comparing the demographics of spiral arm number and a new method for correcting redshift bias}
\author[Hart et al.]{Ross~E.~Hart,$^1$\thanks{E-mail: ross.hart@nottingham.ac.uk} Steven~P.~Bamford,$^1$ Kyle~W.~Willett,$^{2,3}$
Karen~L.~Masters,$^4$\newauthor Carolin Cardamone,$^5$ Chris~J.~Lintott,$^6$ Robert~J.~Mackay,$^1$ Robert~C.~Nichol,$^4$\newauthor
Christopher~K.~Rosslowe,$^1$ Brooke~D.~Simmons,$^7$ Rebecca~J.~Smethurst$^6$
\smallskip\\
$^{1}$School of Physics \& Astronomy, The University of Nottingham, University Park, Nottingham, NG7 2RD, UK\\
$^{2}$School of Physics and Astronomy, University of Minnesota, 116 Church St SE, Minneapolis, MN 55455, USA\\
$^{3}$Department of Physics and Astronomy, University of Kentucky, 505 Rose St., Lexington, KY 40506, USA\\
$^{4}$Institute for Cosmology and Gravitation, University of Portsmouth, Dennis Sciama Building, Burnaby Road, Portsmouth\\
PO1 3FX, UK
$^{5}$Department of Math \& Science, Wheelock College, 200 Riverway, Boston, MA 02215, USA\\
$^{6}$Oxford Astrophysics, Department of Physics, University of Oxford, Denys Wilkinson Building, Keble Road, Oxford OX1 3RH, UK\\
$^{7}$Center for Astrophysics and Space Sciences (CASS), Department of Physics, University of California, San Diego, CA 92093, USA\\
}

\maketitle
\begin{abstract}
The majority of galaxies in the local Universe exhibit spiral structure with a variety of forms. Many galaxies possess two prominent spiral arms, some have more, while others display a many-armed flocculent appearance. Spiral arms are associated with enhanced gas content and star-formation in the disks of low-redshift galaxies, so are important in the understanding of star-formation in the local universe. As both the visual appearance of spiral structure, and the mechanisms responsible for it vary from galaxy to galaxy, a reliable method for defining spiral samples with different visual morphologies is required. In this paper, we develop a new debiasing method to reliably correct for redshift-dependent bias in Galaxy Zoo 2, and release the new set of debiased classifications. Using these, a luminosity-limited sample of $\sim$18,000 Sloan Digital Sky Survey spiral galaxies is defined, which are then further sub-categorised by spiral arm number. In order to explore how different spiral galaxies form, the demographics of spiral galaxies with different spiral arm numbers are compared. It is found that whilst all spiral galaxies occupy similar ranges of stellar mass and environment, many-armed galaxies display much bluer colours than their two-armed counterparts. We conclude that two-armed structure is ubiquitous in star-forming disks, whereas many-armed spiral structure appears to be a short-lived phase, associated with more recent, stochastic star-formation activity.
\end{abstract}


\section{Introduction}
\label{sec:intro}
Spiral galaxies are the most common type of galaxy in the local Universe, with as many as two-thirds of low-redshift galaxies exhibiting disks with spiral structure \citep{Lintott_11,Willett_13,Nair_10,Kelvin_14b}. As star-formation is enhanced in gas-rich disk galaxies \citep{Kennicutt_89,Schmidt_59,Kelvin_14b} understanding spiral structure holds the key to understanding star-formation in the local Universe, yet formulating a single theory to account for all spiral structure still remains elusive.The main theories for the occurrence of spiral arm features in local galaxies initially focused on the idea of being caused by density waves in their disks \citep{Lindblad_63,Lin_64}, but have since been superseded by theories that consider the effects of gravity and disk dynamics \citep{Toomre_81,Sellwood_84}, with most of the work to advance the field of spiral structure theory driven by simulation (eg. \citep{Dobbs_14} and references therein, and discussed further in Sec.4). Using observational studies to test these theories remains a challenge, as visual classifications of both the presence of spiral structure and details of its features are required, which are difficult to obtain when considering the large samples provided by galaxy survey data.

An approach that has been successfully employed to visually classify galaxies in large surveys is citizen science, which asks many volunteers to morphologically classify galaxies rather than relying on a small number of experts. Sophisticated automated methods have also been developed for this purpose, (eg. \citealt{Huertas_11,Davis_14,Dieleman_15}). However, these methods cannot currently completely reproduce the results of visual classifications, particularly in low signal-to-noise images. They also require training sets, meaning that `by eye' inspection methods are still a a requirement. Galaxy Zoo 1 \citep[GZ1;][]{Lintott_08,Lintott_11} was the first project to collect visual morphologies using citizen science, by classifying galaxies from the Sloan Digital Sky Survey (SDSS) as either `elliptical' or `spiral'. Using this method, each galaxy is classified by several individuals, and a likelihood or `vote fraction' of each galaxy having a particular feature is assigned as the fraction of classifiers who saw that feature. GZ1 classifications collected in this way have been used to compare galaxy morphology with respect to colour \citep{Bamford_09,Masters_10b,Masters_10a}, environment \citep{Bamford_09,Skibba_09,Darg_10a,Darg_10b}, and star formation properties \citep{Tojeiro_13,Schawinski_14,Smethurst_15}. 

Following from the success of GZ1, more detailed visual classifications were sought, including the presence of bars, and spiral arm winding and multiplicity properties. Thus, Galaxy Zoo 2 (GZ2) was created \citep[][hereafter W13]{Willett_13}, in which volunteers were asked more questions about a subsample of GZ1 SDSS galaxies. The main difference between GZ2 and GZ1 was that visual classifications were collected using a `question tree' in GZ2, to gain a more exhaustive set of morphological information for each galaxy. GZ2 has already been used to compare the properties of spiral galaxies with or without bars \citep{Masters_11,Masters_12,Cheung_13}, look for interacting galaxies \citep{Casteels_13}, as well as looking for relationships between spiral arm structure and star formation \citep{Willett_15}. This `question tree' method has since been used in a similar way to measure the presence of detailed morphological features in higher redshift galaxy surveys (eg. \citet{Melvin_14,Simmons_14}), and other \textsc{Zooniverse}\footnote{https://www.zooniverse.org/} citizen science projects. 

An issue that arises in both visual and automated methods of morphological classification is that detailed features are more difficult to observe in lower signal-to-noise images (ie. observed from a greater distance). In Galaxy Zoo, this has been termed as classification bias. It is imperative that classification bias is removed from morphological data, as it leads to sample contamination from galaxies being incorrectly assigned to some categories. This means that any observational differences between samples can be significantly reduced.

Classification bias manifested itself in GZ1 with galaxies at higher redshift having lower `spiral' vote fractions, which were corrected using a statistical method \citep{Bamford_09}. The application of a question tree in GZ2 to look for more detailed features means that correcting for biases is more complicated than in GZ1. In particular, there are questions with several possible answers, and debiasing one answer with respect to each of the others is therefore a more difficult process for GZ2.

The paper is organised as follows. In Sec.~\ref{sec:data}, the sample selection and galaxy data are described. In Sec.~\ref{sec:redshift_bias}, we describe a new debiasing method that has been created to account for the classification bias in the GZ2 questions with multiple possible answers. In Sec.~\ref{sec:results}, samples of GZ2 spiral galaxies are defined and sorted by arm multiplicity. This is a case where the new debiasing method is required as there are multiple responses to that question. After reviewing relevant theoretical and observational literature, we examine the demographics of spiral galaxies with respect to arm multiplicity, and begin to explore the processes that influence the formation and evolution of spiral arms in Sec.~\ref{sec:results}. The results are summarised in Sec.~\ref{sec:conclusions}.

This paper assumes a flat cosmology with $\Omega_\mathrm{m} = 0.3$ and $H_0 = 70\;\mathrm{km\,s^{-1}\,Mpc^{-1}}$.
\section{Data}
\label{sec:data}
\subsection{Galaxy properties and sample selection}
\label{sec:sample}

\begin{figure}
		\centering
		\includegraphics[width=0.45\textwidth]{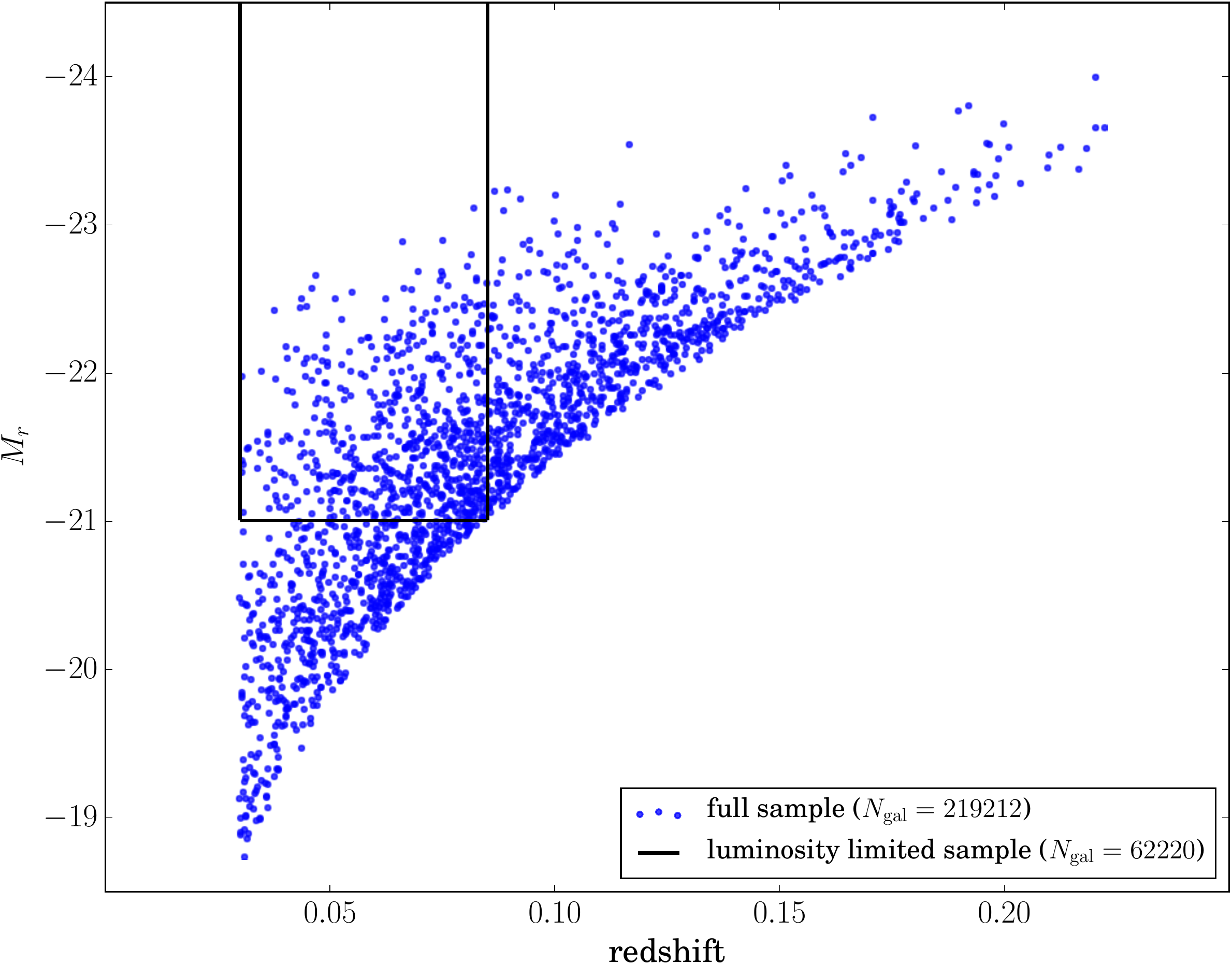}
    \caption{The $r$-band luminosity versus redshift distribution of our \textit{full sample} (blue points), with the region enclosing our $0.03<z<0.085$, $M_r  \leq -21$ \textit{luminosity-limited sample} indicated by black lines.
		\label{fig:vl_sample}}
\end{figure}

We make use of morphological information from the public data release of Galaxy Zoo 2. The galaxies classified by GZ2 were taken from the Sloan Digital Sky Survey (SDSS) Data Release 7 (DR7; \citealt{Abazijian_09}). The SDSS main galaxy sample (MGS) is an $r$-band selected sample of galaxies in the legacy imaging area targeted for spectroscopic follow-up \citep{Strauss_02} The GZ2 sample contains essentially all well-resolved galaxies in DR7 down to a limiting absolute magnitude of $m_r \leq 17$, supplemented by additional sets of galaxies in Stripe 82 for which deeper, co-added imaging exists (see W13 for details).  In this paper we only consider galaxies with $m_r \leq 17$ that were classified in normal-depth SDSS imaging and which have DR7 spectroscopic redshifts. We refer to this as our \textit{full sample}, containing $228,201$ galaxies, to which the debiasing procedure described in Sec.~\ref{sec:new_method} is applied. We require redshifts in order to correct the sample for a distance-dependent bias, as described in Sec.~\ref{sec:biases}.

Petrosian aperture photometry in $ugriz$ filters is obtained from the SDSS DR7 catalogue. Rest-frame absolute magnitudes corrected for Galactic extinction are those computed by \citet{Bamford_09}, using \textsc{kcorrect} \citep{Blanton_07}. Galaxy stellar masses are determined from the $r$-band luminosity and $u-r$ colour using the calibration adopted by \citet{Baldry_06}.

In order to study galaxy properties in a representative manner in Sec.~\ref{sec:results}, we define a \textit{luminosity-limited sample} with $0.03<z<0.085$ and $M_r \le -21$, containing $62,220$ galaxies. The luminosity versus redshift distribution of our \textit{full sample}, and the limits of our \textit{luminosity-limited sample}, are shown in Fig.~\ref{fig:vl_sample}.  These limits approximately maximize the sample size, given the $m_r \le 17$ limit on the \textit{full sample}. The lower redshift limit avoids a small number of galaxies with very large angular sizes, and hence accompanying morphological, photometric and spectroscopic complications. The upper redshift limit also corresponds to that for which we have reliable galaxy environmental density data from \cite{Baldry_06}, which we will make use of in this paper.  

The \textit{luminosity-limited sample} is incomplete for the reddest galaxies at $\log (M/M_{\odot}) < 10.6$ (calculated using the method in \citealt{Bamford_09}). Where necessary we therefore consider a \textit{stellar mass-limited sample} of 41,801 galaxies, created by applying a limit of $\log (M/M_{\odot}) \geq 10.6$ to the \textit{luminosity-limited sample}.
\subsection{Stellar population models}
\label{sec:SEDs}

In Sec.~\ref{sec:colours}, we evaluate potential star-formation histories by comparing observed galaxy colours. Spectral energy distributions (SEDs) are derived from \citet{BC_03}, for a range of ages and SFHs using the initial mass function from \citet{Chabrier_03}. For star-forming galaxies in the SDSS, the mean stellar metallicity varies from $Z \approx 0.7 Z_{\odot}$ for $M \sim 10^{10.6} M_\odot$ (the lower limit of the stellar mass-limited sample) to $Z \approx Z_{\odot}$ for $M \sim 10^{11} M_\odot$ \citep{Peng_15}. As we expect most spirals to be blue star-forming galaxies (eg. \citealt{Bamford_09}), we approximate the metallicity of the stellar mass-limited spiral sample using a metallicity value of $Z=Z_\odot$. Two dust extinction magnitudes of $A_{V}$=0 and $A_{V}$=0.4 are considered \citep{Calzetti_00}. Equivalent colours for each of the star formation and dust extinction models are calculated for each of the SDSS $ugriz$ filters \citet{Doi_10}. Full details of how the models are derived can be found in \citet{Duncan_14}.
\subsection{Quantifying morphology with Galaxy Zoo}
\label{sec:gz_morphologies}

\begin{figure*}
		\centering
		\includegraphics[width=1\textwidth]{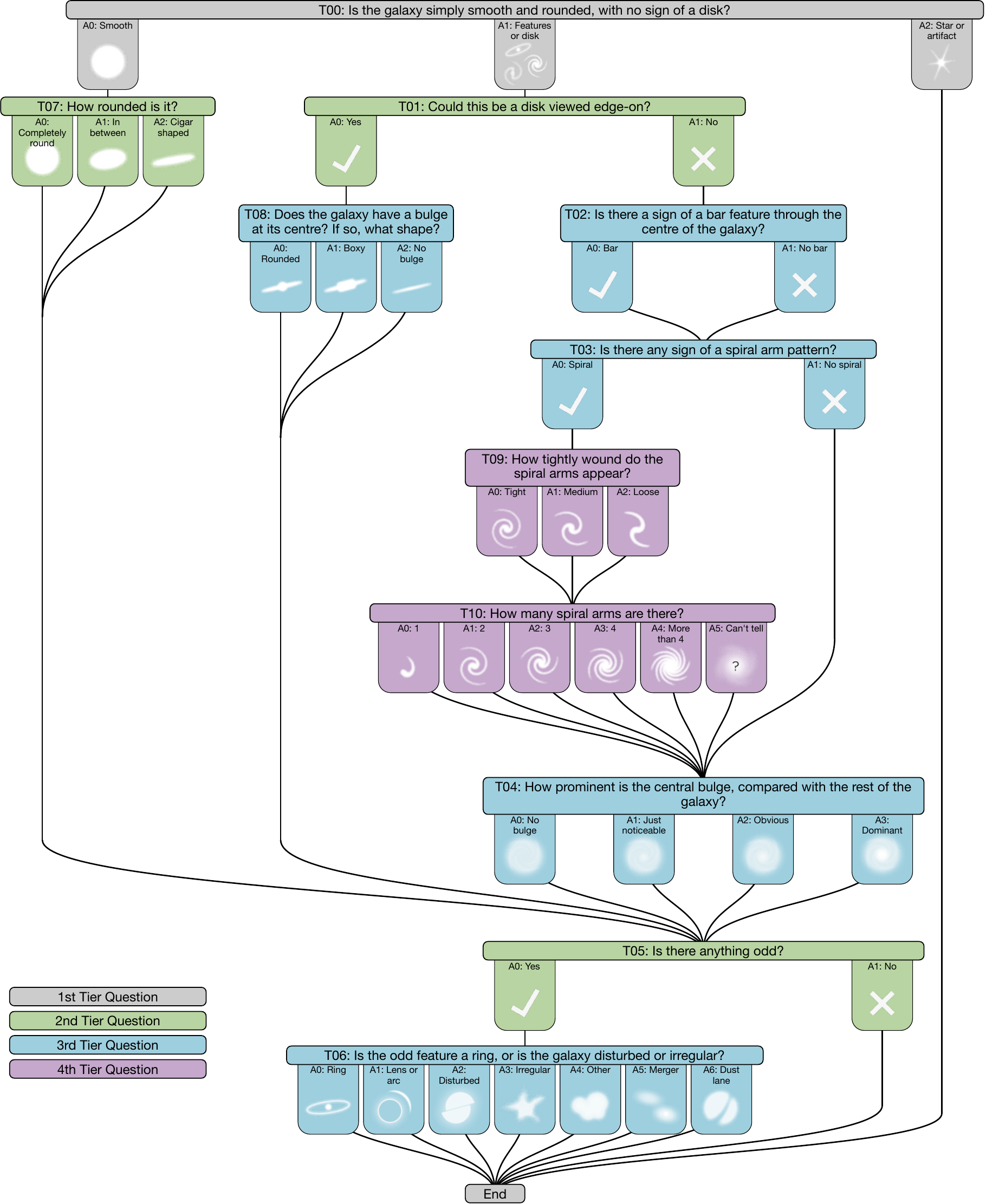}
    \caption{Diagram of the question tree used to classify galaxies in GZ2. The tasks are colour-coded by their depth in the question tree. As an example, the arm number question (T10) is a fourth-tier question --- to answer that particular question about a given galaxy, a participant needs to have given a particular response to three previous questions (that the galaxy had features, was not edge-on and had spiral arms).}
    \label{fig:question_tree}
\end{figure*}

In GZ2, morphological information for each galaxy was obtained by asking participants to answer a series of questions. The structure of this question tree is shown in Fig.~\ref{fig:question_tree}. Typically, each image was viewed by $\ga 40$ people (W13), although no user will explicitly answer every question in the tree for a particular galaxy. To reach the questions further down the tree, it is required that another question has been answered with a specific response. For each question, the responses are each represented by the `vote fraction` $p$ assigned to each possible answer. For any given question, the sum of the vote fractions for all possible answers adds up to one. Considering the `edge-on' question (T01 in Fig.~\ref{fig:question_tree}), a classifier would only answer that question if they answered `features/disk' for T00. For example; if a galaxy was classified by 40 people, and 30 of those said they saw features, whilst the other 10 claimed it was smooth, then the corresponding vote fractions are $p_{\mathrm{features}}=0.75$ and $p_\mathrm{{smooth}}=0.25$. Only the 30 classifiers who saw `features' would then answer the `edge-on' question (T11 of Fig.~\ref{fig:question_tree}). If 15 of those said the galaxy was edge-on, and 15 said it was not, the corresponding vote fractions would be $p_\mathrm{edge-on}=0.5$ and $p_\mathrm{not \, edge-on}=0.5$. 

In order to reduce the influence of unreliable classifiers, W13 downweighted individual volunteers who had poor agreement with the other classifiers.  Throughout this paper we refer to these weighted vote fractions as the `raw' quantities. Before using these GZ2 vote fractions to study the galaxy population, we must first consider the issue of classification bias, as we shall in Sec.~\ref{sec:biases}.

Traditional morphologies assign each galaxy to a specific class, usually determined by one, or occasionally a few, experts. In contrast, Galaxy Zoo provides a large number of independent opinions on specific morphological features for each galaxy.  This allows us to consider both the inherent `subjectiveness' and observational uncertainties of galaxy morphology, and hence control the compromise between sample contamination and completeness.

There are two principal ways in which galaxy morphologies can be quantified using Galaxy Zoo vote fractions. The first is to consider means of the vote fractions over specific samples or bins divided by some other property.  These average vote fractions can then be used to study variations in the morphological content of the galaxy population.  Individual galaxies are not given specific classifications.  There is no population of `unclassified', and hence ignored, galaxies.  This approach has been taken by \citet{Bamford_09}, \citet{Casteels_13}, \citet{Willett_15}, and various other studies. With this method, the vote fractions of all galaxies can be considered together; even galaxies with a small (but non-zero) vote fraction for a given property count towards the statistics. Effectively, this approach considers the vote fractions as an estimate of the probability of a galaxy belonging to a particular class.

The second approach is to divide the galaxy sample into different morphological categories, either by applying a threshold on the vote fractions, or choosing the class with the largest vote fraction. Such methods have been used by \citet{Land_08}, \citet{Skibba_09}, \citet{Galloway_15} and many more.  One advantage of this approach is that each galaxy is assigned to a definite class, with the threshold tuned to ensure a desired level of classification certainty.  However, a set of `uncertain' or `unclassified' galaxies may remain.  In some analyses these will require special attention.

These different approaches are also relevant for how questions at different levels in the tree are combined.  For example, a participant is only asked if they can see spiral arms when they have already answered that they can see features in the galaxy and that the galaxy is not an edge-on disc.  The vote fraction for spiral arms therefore represents the conditional probability of spiral arms \emph{given that} features are discernible \emph{and} that the galaxy is not edge-on. When considering whether a galaxy displays spiral arms, one should account for the answers to these previous questions in the tree.  One can treat vote fractions as probabilities, multiplying them to obtain a `probability' that a galaxy displays any features, is not edge-on and possesses spiral arms.  Alternatively, one may select a set of galaxies that display features, are not edge-on and possess spiral arms, by applying some thresholds to the vote fractions for each question in turn. (See \citet{Casteels_13} for a more thorough discussion of these issues.)

The primary morphological feature we will focus on in this paper is the apparent number of spiral arms displayed by a galaxy.  Some of the classes for this feature, though, contain a relatively low fraction of the total spiral population.  In addition, the vote fractions for the preferred answer are often fairly low, with votes distributed over several answers.  In such cases, averaging the vote fractions over the full sample does not work particularly well, as noise from more common galaxy classes overwhelms the subtle signal from rarer classes.  In this paper we therefore prefer to assign galaxies to morphological samples by applying a threshold or taking the answer with the largest vote fraction.

\section{Correcting for redshift-dependent classification bias}
\label{sec:redshift_bias}
\subsection{Biases in the Galaxy Zoo sample}
\label{sec:biases}

Galaxies of a given size and luminosity appear fainter and smaller in the SDSS images if they are at higher redshifts. To correct for this, galaxy images in GZ2 are scaled by Petrosian radius (W13). As this means that galaxies at further distances are scaled to have the same angular size, their pixel resolution is lower. Detailed features can therefore be more difficult to distinguish in galaxies at higher redshift. As a result, visual galaxy classifications are biased, as fewer galaxies are classified as having the more detailed features at higher redshift, making a sample of galaxies with the these features incomplete.

It should be noted that such biases are not exclusive to Galaxy Zoo. Difficulty in detecting faint features in lower signal-to-noise galaxies is an inherent property of any visual or automated method of galaxy classification. The advantage of using Galaxy Zoo classifications is that they give a statistical method of measuring galaxy morphology. As each of the galaxies in the \textit{full sample} has been visually classified by a number of independent observers, the apparent evolution in the presence of features can be modelled, and biases corrected accordingly.

Incompleteness and contamination are defects that arise in a sample where an inherent redshift bias affects the classifications. Incompleteness affects the `harder to see' features: the vote fractions for these features decrease with redshift, leaving us with poor number statistics for a sample we wish to define as having that feature.  Contamination is the converse effect that appears in the `easier to see' categories.  For these responses, the vote fractions decrease with redshift, meaning that any samples defined using the Galaxy Zoo classifications will also include mis-classified galaxies that should have actually been included in one of the `harder to see' categories. Any intrinsic differences between samples that one wishes to compare may therefore be negated.

\begin{figure}
		\centering

        \includegraphics[width=0.45\textwidth]{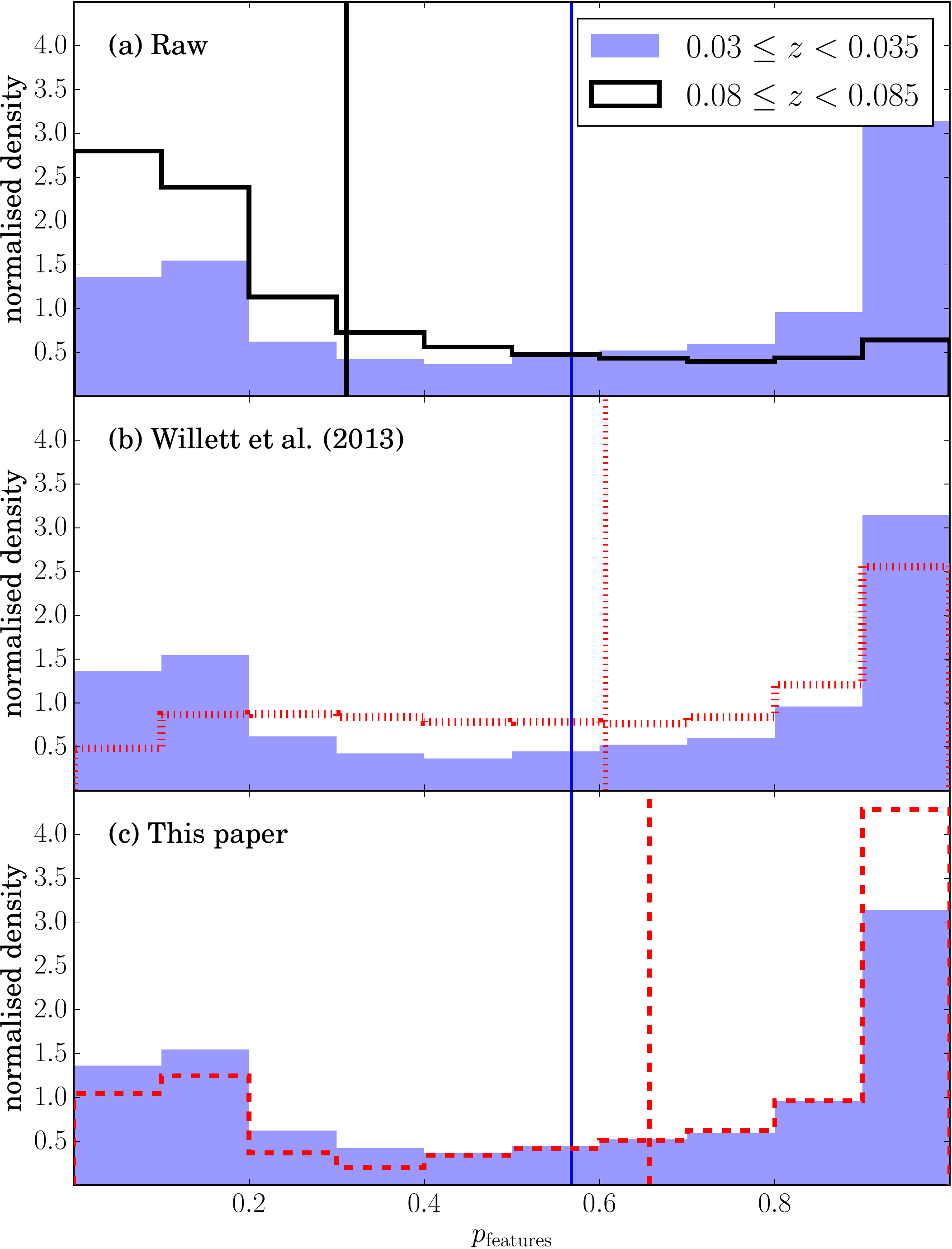}

        \caption{Histograms of vote fractions for the `features' response to the `smooth or features' question in GZ2. In each of the panels, the blue filled histogram shows the raw vote distribution for a low-redshift $0.03 \leq z < 0.035$ slice of the \textit{luminosity limited sample}. The unfilled histograms show the equivalent distribution for a higher-redshift $0.08 < z \leq 0.085$ sample. The vertical lines show the mean vote fractions.}

        \label{fig:vote_histograms}

\end{figure}

The effect of redshift bias is shown in Fig.~\ref{fig:vote_histograms}a, where the answer to the `smooth or features' question is compared for high and low-redshift samples. The redshift range of the SDSS sample is shallow enough to argue that there should be minimal change in the overall population of galaxies \citep{Bamford_09,Willett_13}. In a \textit{luminosity-limited sample}, the level of completeness should also be the same at all redshifts, meaning that the overall populations of the high and low redshift samples should be equivalent. However, Fig.~\ref{fig:vote_histograms}a shows that the higher redshift vote fractions are dramatically skewed to lower values- generally, people are having greater difficulty in detecting the presence of features in the higher redshift images. Thus, there are fewer votes for galaxies showing `features' and consequently more votes for galaxies being `smooth'. If one wished to compare a sample of galaxies with `features' against one that is `smooth' using the raw vote fractions, the number of galaxies with `features' would be incomplete and the `smooth' sample would be contaminated.

\subsection{Previous corrections for redshift bias in GZ2}
\label{sec:previous_method}

The previous debiasing procedure applied to both GZ1 and GZ2 focused on correcting the vote fractions of the galaxy samples by adjusting the mean vote fractions as a function of redshift. The method was first proposed in \cite{Bamford_09}, and updated for GZ2 in W13. The method successfully adjusts the mean vote fractions for questions with two dominant answers, as can be seen from the vertical lines in Fig.~\ref{fig:vote_histograms}b: the mean of the debiased high-redshift sample is much closer to the mean of the low-redshift sample than for raw vote distributions (Fig.~\ref{fig:vote_histograms}a).

However, this technique has two limitations that make it unsuitable if we want to divide a galaxy sample into different morphology subsets.  The first issue is that adjustment of the mean vote fraction does not necessarily lead to correct adjustment of individual vote fractions. This can be seen in Fig.~\ref{fig:vote_histograms}b.  Although the mean vote fraction for the high-redshift sample has been correctly adjusted to approximately match the low-redshift sample, the overall distribution does not. There is an excess of debiased votes in the middle of the distribution, and fewer votes for the tails of the distribution at $p \approx 0$ and $p \approx 1$. This effect is important if we wish to divide our sample into different subsets by morphological type. As the shape of the histograms is not consistent with redshift, the fraction of galaxies with $p_\mathrm{features}$ greater than a given threshold can also vary with redshift. 

\begin{figure}
		\centering

        \includegraphics[width=0.45\textwidth]{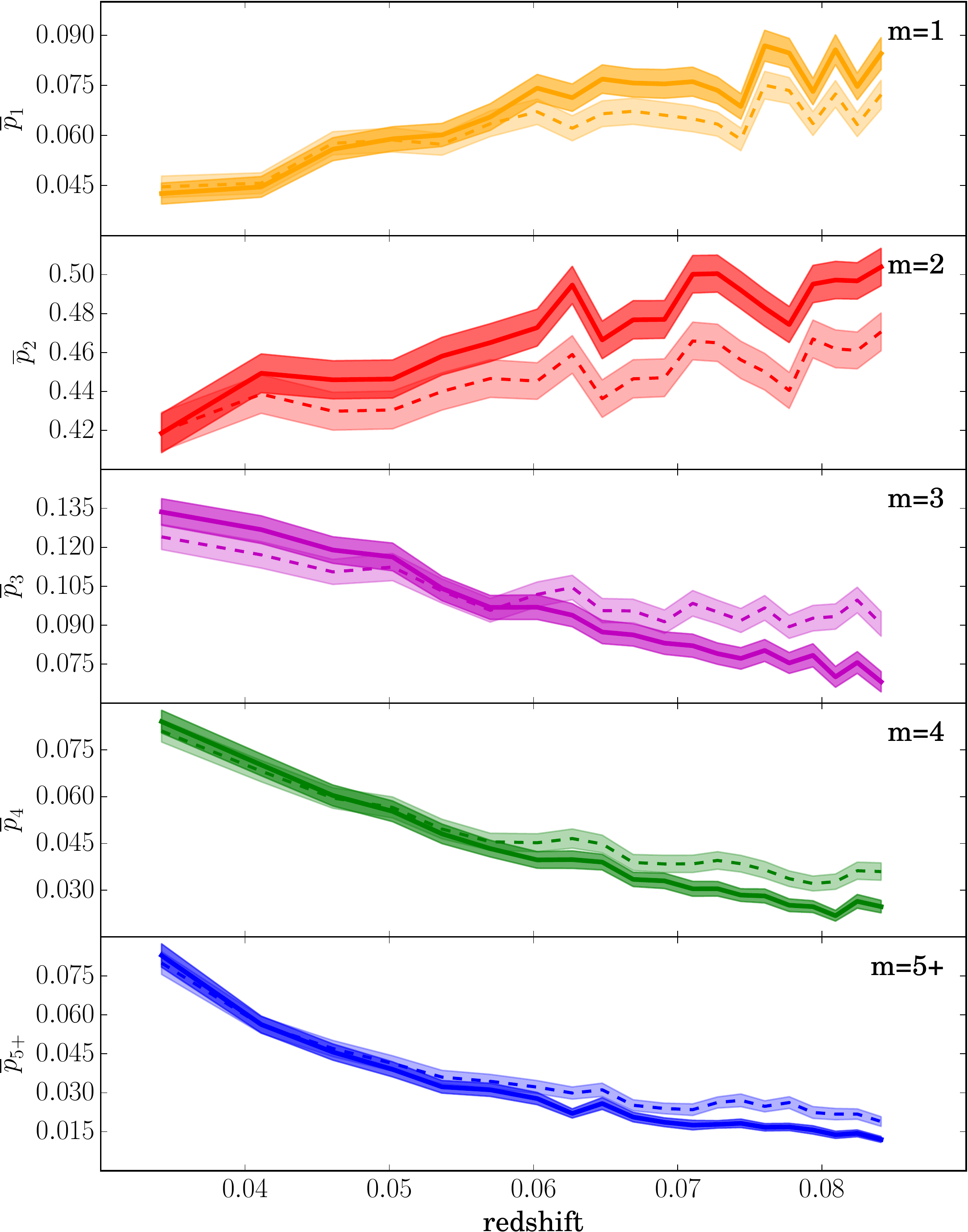}

        \caption{Mean vote fractions for each of the arm number responses to the `arm number' question (T10 in Fig.~\ref{fig:question_tree}. The sample consists of galaxies from the \textit{luminosity-limited sample}, with $p_\mathrm{features} \times p_\mathrm{not \, edge-on} \times p_\mathrm{spiral} > 0.5$ (with vote fractions taken from the W13 debiased catalogue). The solid lines show the mean arm number vote fractions obtained using the raw vote classifications, and the dashed lines indicate the same quantity obtained using the W13 debiased values. The shaded regions indicate the $1 \sigma$ error on the mean.}

        \label{fig:arm_bias}

\end{figure}

As described in section \ref{sec:gz_morphologies}, GZ2 utilises multiple answered questions to obtain more detailed classifications than GZ1. In cases where the votes are split between multiple categories, the debiasing method from W13 does not always adjust the vote fractions correctly. We show this effect for the `spiral arm number' question (T10 of Fig.~\ref{fig:question_tree}), in Fig.~\ref{fig:arm_bias}. A sample of `secure' spiral galaxies with $p_{\mathrm{features}} \times p_{\mathrm{not \, edge-on}} \times p_{\mathrm{spiral}} > 0.5$ is selected, (with the vote fractions corresponding to the debiased values from W13), and plot the mean vote fractions with respect to redshift for each of the arm number responses. A clear trend in $p_\mathrm{arm \, number}$ is observed: the mean vote fractions vary systematically with redshift, even after the W13 correction has been applied. For this question, the answers with more spiral arms (3, 4, or 5+ spiral arms) are the `harder to see' features meaning that there are fewer votes for these categories at higher redshift, which instead increase the 1 and 2 arm vote fractions. The 3, 4 and 5+ spiral arm samples of spiral galaxies therefore suffer from incompleteness. This is of particular importance in this case for two reasons. Firstly, as this is a `fourth order' question, as can be seen in Fig.~\ref{fig:question_tree}, then the sample size is limited, as three questions must have been answered `correctly' previously for a galaxy to be classified as spiral. Secondly, the 3, 4 and 5+ arm responses have low mean vote fractions overall, of $\lesssim 0.1$. Thus, the number statistics for these categories are very low, meaning they will suffer from high levels of noise. Correspondingly, the 1 and 2 armed spiral samples would suffer from contamination from galaxies that should have been classified as 3, 4 or 5+ armed.

\subsection{A new method for removing redshift bias}
\label{sec:new_method}

Given the limitations described in Sec.~\ref{sec:previous_method}, we attempt to construct a new method of debiasing the GZ2 data more effectively. When considering a question further down the question tree with low number statistics, such as the spiral arm question, we prefer to use a thresholding technique rather than using the weighted vote fractions (see Sec.~\ref{sec:gz_morphologies} for a descriptions of both methods). Using the arm number question as an example, the `2 spiral arms' response dominates the overall vote fractions, making up $\sim 60 \%$ of the votes, as can be seen in Fig.~\ref{fig:arm_bias}. The rarer responses of 3, 4 or 5+ arms have much lower number statistics overall, with only $\sim 10 \%$ of the votes. The mean values can therefore be affected by the noise in the dominant category, which will be much larger than the noise for the rarer category. We therefore divide our galaxy sample into different sub-samples when comparing galaxies by spiral arm number.

Unlike the debiasing method in W13, our new method aims to make the vote distributions themselves as consistent as possible rather than purely aiming for consistency in the mean vote fraction values. As each galaxy is classified by 40 or more volunteers (W13), we have enough data to model the evolution of the vote distributions as a function of redshift. Different classifiers will have different sensitivity to picking out the most detailed features. Thus, as samples at higher redshift are considered, and hence with poorer image quality, we expect the vote fraction distributions to also evolve as some classifiers become less able to see the most detailed features. We aim to account for this bias by modelling the vote fraction distributions as a function of redshift, and correcting the higher redshift vote distributions to be as similar as possible to equivalent vote distributions at low redshift. 

We first define samples of galaxies for each of the questions in turn. The sample is then binned in terms of the intrinsic galaxy properties of size and luminosity, and each of these bins is divided into redshift slices. We then attempt to model the vote distributions for each of the bins with respect to redshift, and thus match their distributions to those at low redshift. This means that if a vote fraction threshold is applied, the fraction of galaxies with a given feature remains constant: at each redshift, the sample is composed of the galaxies that are most likely to have that particular feature. 

It must be noted that such a method could still be limited by small-number statistics, which is particularly common at higher redshifts. In the case that a feature's vote fraction drops to 0, we can not `add' votes for a feature --- it is only possible to debias the galaxies with $p>0$, where there is evidence for a feature being present. This remains a problem for the categories where the vote fractions are lowest, such as in the responses to the odd feature question (T06 in Fig.~\ref{fig:question_tree}).
\subsubsection{Sample selection for each question}
\label{sec:sample_selection_per_question}

As GZ2 morphologies are classified with a decision tree (see section \ref{sec:gz_morphologies}), not all of the questions were answered by each of the volunteers for a given galaxy. Answering the spiral arm number question is not appropriate for all of the galaxies in the sample: if a galaxy has no spiral features, yet a volunteer answered the spiral arm question, that galaxy would contribute `noise' to the answers to that question. To avoid `noise' introduced by incorrectly classified galaxies, clean galaxy samples are defined with $p > 0.5$. For the first question, this corresponds to all of the galaxies, as each classifier answered that particular question for each galaxy. However, when questions further down the tree are considered, this is not the case. The equivalent $p>0.5$ for the spiral arm question would only include the galaxies with $p_{\textrm{features}} \times p_{\textrm{not edge-on}} \times p_{\textrm{spiral}} > 0.5$. 

For each of the questions in turn, we define a sample of galaxies with which we will apply the new debiasing procedure. These samples are defined using a cut of $p>0.5$ (corresponding to $p_{\textrm{features}} \times p_{\textrm{not edge-on}} \times p_{\textrm{spiral}} > 0.5$ for the spiral arm question for example). A further cut of $N \geq 5$ (where $N$ is the number of classifications) is also imposed to ensure that each galaxy has been classified by a significant number of people to reduce the effects of Poisson noise. In this case, the vote fractions must be the debiased vote values, to ensure each sample is as complete as possible (see Sec.~\ref{sec:biases}) as we look at each question. The order in which the questions are debiased is important: to define a complete sample of galaxies to be used for the debiasing of a particular question, all questions further up the question tree must have been debiased beforehand.
\subsubsection{Binning the data}
\label{sec:binning}

\begin{figure}
		\centering

        \includegraphics[width=0.45\textwidth]{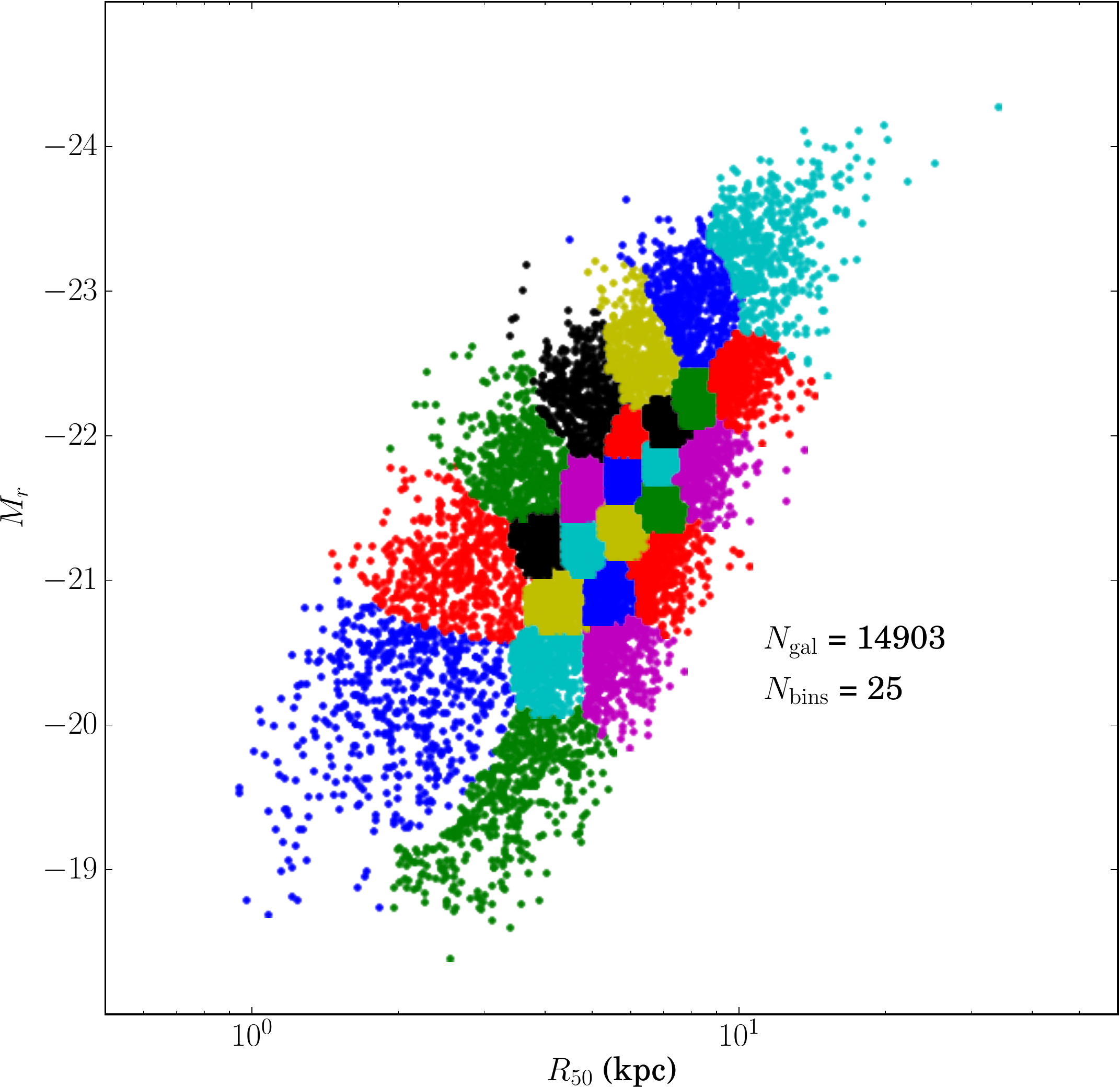}

        \caption{Voronoi bins for the more than 4 arms (A4) answer to the spiral arm number question (T10). The sample is defined using the method described in Sec.\ref{sec:binning}, and binned in terms of $\log R_{50}$ and $M_r$. Different bins are defined with different colours. Each Voronoi bin is further subdivided into several redshift bins.}

        \label{fig:voronoi_bins}

\end{figure}

It is expected that the ability to discern the presence of a particular feature will depend on intrinsic galaxy properties.  For example, larger, brighter galaxies may be easier to classify over a wider redshift range. Conversely, fainter galaxies may show stronger features, as both overall galaxy morphology \citep{Maller_08,Bamford_09} and spiral arm morphology \citep{Kendall_15} have stellar mass dependences. To account for these possible variations, we bin the data in terms of $M_r$ and $\log(R_{50})$ for each answer in turn. We use the \texttt{voronoi\_2d\_binning} package from \cite{Cappellari_03},to ensure the bins will have an approximately equal number of galaxies. Fig.~\ref{fig:voronoi_bins} shows an example of the Voronoi~binning for the 5+ arms response to the arm number question. When Voronoi~binning the data for each of the answers, only the $N_{\mathrm{gal}}$ galaxies with $p>0$ are included, meaning that the `signal' of galaxies is evened out over all of the Voronoi bins. We aim to have $\sim 30$ Voronoi bins for each of the questions, so the desired number of galaxies in each bin is given by $N_{\mathrm{gal}}/30$.

After Voronoi~binning the data in terms of their intrinsic properties of size and brightness, we further divide each bin into redshift bins, to allow us to study how the vote distributions change with redshift. Each redshift bin is constrained to contain $\geq 50$ galaxies. This binned data is used for the debiasing methods described in the next section.
\subsubsection{Modelling redshift bias}
\label{sec:debiasing}

For each of the possible responses to each question, a method is applied to correct for the redshift bias in the sample, aiming to make the vote distributions for each answer consistent with redshift. The two methods that we employ to achieve this are described below. 

\begin{figure}
		\centering

        \includegraphics[width=0.45\textwidth]{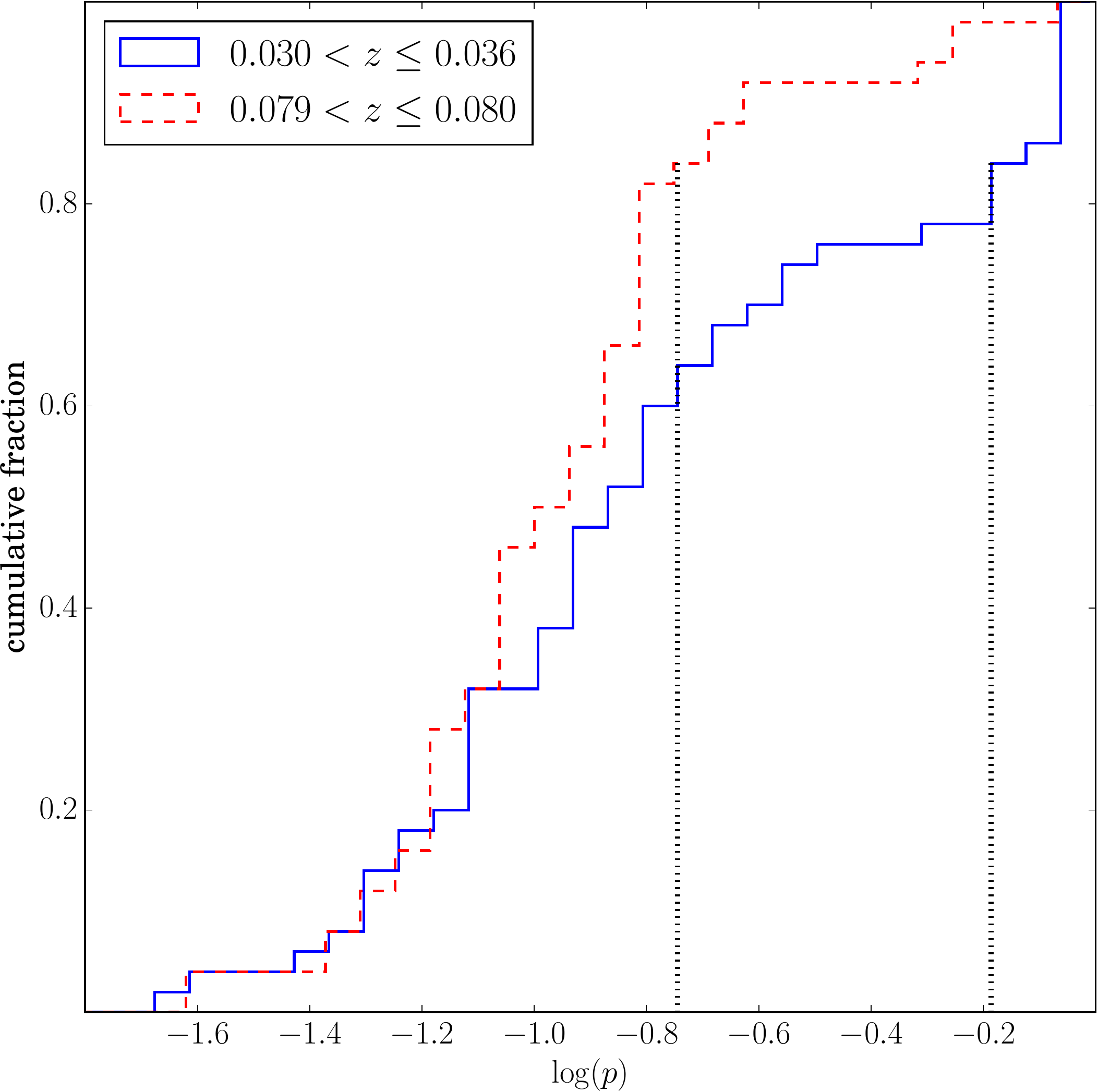}

        \caption{An example of vote distributions for an example Voronoi bin for the `features or disk' answer to the `smooth or features' question. Each of the galaxies in the high-redshift bin (red dashed line) is matched to its closest equivalent low-redshift galaxy (blue solid line) in terms of cumulative fraction. The dotted lines indicate the `matched' values for an example galaxy with $\log(p) \approx -0.8$, and an equivalent low-redshift value of $\log(p) \approx -0.2$ (corresponding to $p_{\mathrm{raw}}=0.18$ and $p_{\mathrm{debiased}}=0.65$). We plot $\log(p)$ on the x-axis rather than $p$ to make the two distributions more easily discernable.}

        \label{fig:histogram_matching}

\end{figure}

The first method we utilise to remove redshift bias simply matches the shapes of the histograms on a bin-by-bin basis. The cumulative distribution for the lowest redshift sample in a given Voronoi bin is used as a reference for how the shape of the histogram would look if it were viewed at low redshift. An example of this method is shown in Fig.~\ref{fig:histogram_matching}, in which the `features or disk' answer to the `smooth or features' question is considered. For both the low redshift bin and the high redshift bin, the vote fractions are ranked in order of low to high. Each of the galaxies in the high redshift bin is then matched to its low redshift equivalent by finding the galaxy with the closest cumulative fraction in the low redshift bin. An example of this technique is shown by the vertical lines of Fig.~\ref{fig:histogram_matching}. In this case, a galaxy with cumulative fraction of $\approx 0.8$ in the high redshift bin has $p_{\mathrm{features}} \approx 0.18$. A galaxy at the same cumulative fraction in the low-redshift bin has $p_{\mathrm{features}} \approx 0.65$, so this is the debiased value assigned to that galaxy. This is repeated for each galaxy and for each of the high redshift bins in turn. Applying a vote fraction threshold for a given response gives the same fraction of the population above that threshold in all of the redshift bins, with the galaxies most likely to have a feature making up the population of galaxies above that threshold. 

\begin{figure*}
		\centering

        \includegraphics[width=0.975\textwidth]{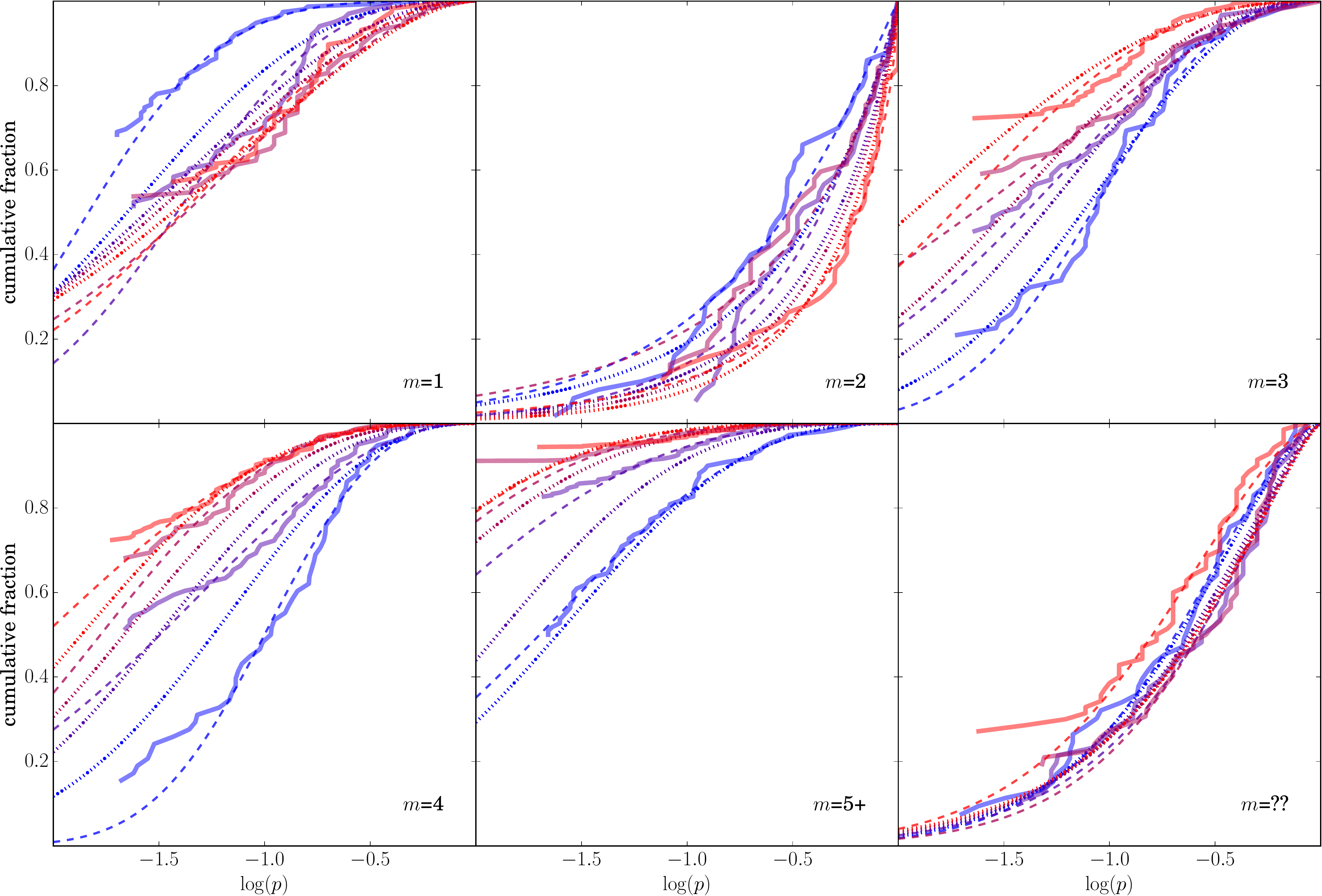}

        \caption{An example of a single Voronoi bin fit for the \textit{arm number} question. The red line indicates the highest redshift bin, and the blue line indicates the lowest redshift bin. The solid lines indicate the raw $p$ histograms, and the dashed lines show the best fit function to each of them. The dotted lines show the corresponding approximation from the continuous fit to the $k$ and $c$ values.}

        \label{fig:function_fit}

\end{figure*}

\begin{figure*}
		\centering

        \includegraphics[width=0.975\textwidth]{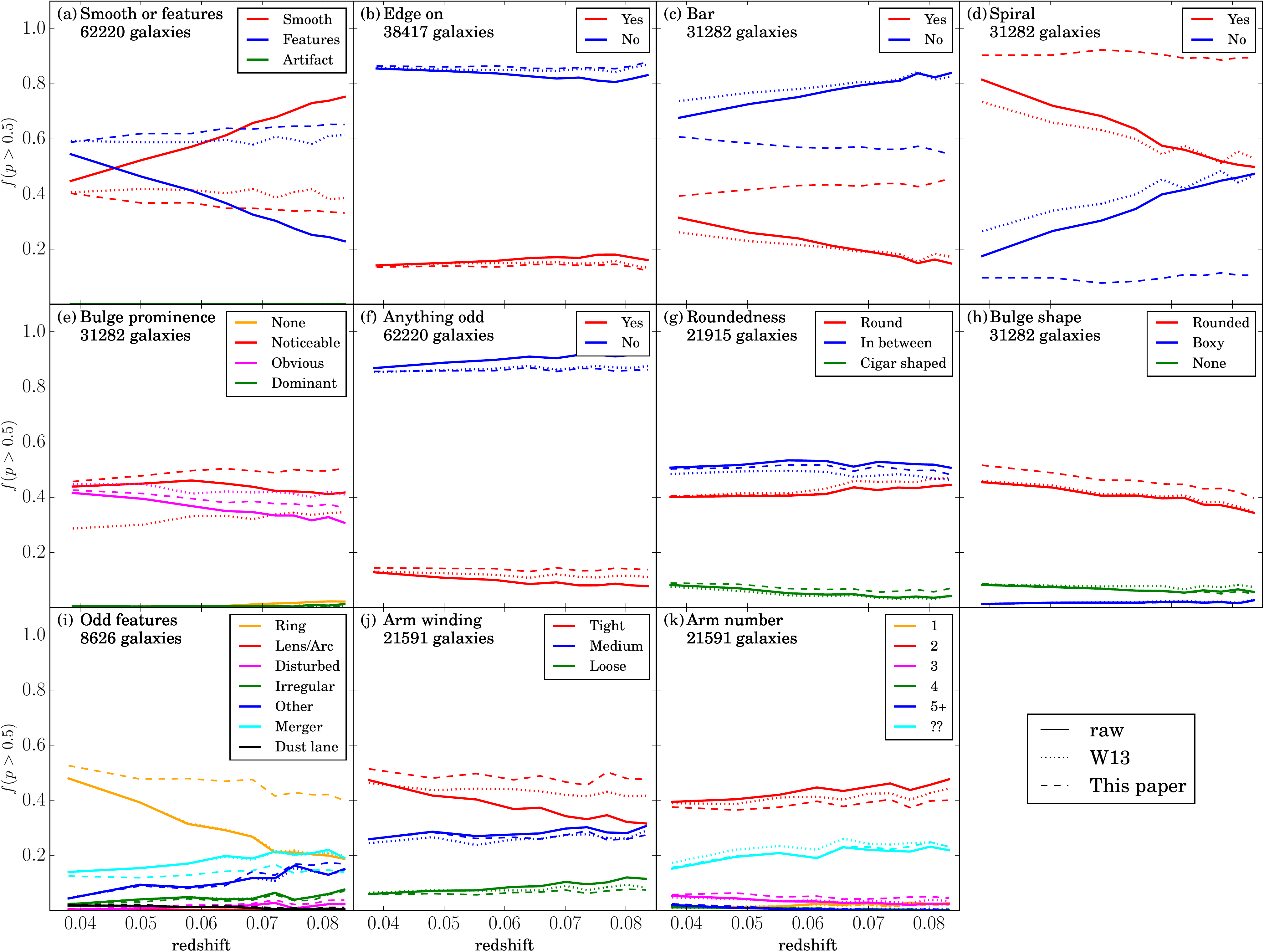}

        \caption{Number of galaxies with $p>0.5$ for each of the questions debiased using the method described in Sec.~\ref{sec:new_method}. The solid lines indicate the raw vote fractions and the dashed lines indicate the debiased vote fractions. The dotted lines indicate the same fractions using the W13 debiasing method. The total sample here is composed of galaxies in the \textit{luminosity-limited sample} with $p>0.5$ (as described in Sec.~\ref{sec:sample_selection_per_question}).}

        \label{fig:all_thresholds}

\end{figure*}

The main strength of this method is that any vote distribution can be modelled in this way, irrespective of the overall shape. However, a potential weakness is that noise can be introduced due to the discretisation of the data. To limit this issue, each redshift bin has a `good' signal of $\geq50$ galaxies. This effectively `blurs' any trends with redshift, and can actually lead to an over-correction of vote fractions, which can be seen in Fig.~\ref{fig:vote_histograms}c. Although the overall histogram shape is well matched when a slice at $0.08 \leq z < 0.085$ is considered, we see too many galaxies with $p \approx 1$ compared to the low redshift data. This issue is purely caused by the discretisation of the individual bins: although the trends can be modelled overall, any trends within individual bins cannot. If there is a redshift trend \emph{within} a bin, then the fraction of galaxies with the more difficult to see features will preferentially reside in the lower redshift ends of the bins. This effect leads to an overestimate of the number of galaxies with the more difficult to see features. Fig.~\ref{fig:all_thresholds}a shows the debiased trends of the `features or disk' question, which was debiased using the `bin-by-bin' method, which shows that the method slightly over-corrects the redshift trend in the number of galaxies classified with $p_{\mathrm{features}} > 0.5.$

One potential solution would be to bin the data more finely. However, there is no `ideal' solution to this problem, as fewer galaxies in each bin would mean that the redshift range that each bin occupies is smaller, but the noise in each of the bins is larger. 

To attempt to remove the discrete nature of the correction in the `bin-by-bin' method, we use an alternative method that models the vote distributions with analytic functions. For each of the redshift bins, we plot a cumulative histogram of $\log(p)$ against the cumulative fraction. Examples of some of these cumulative histograms are plotted as the solid lines in Fig.~\ref{fig:function_fit}. It can be seen that there is a clear evolution in the distributions with redshift. This effect is most prominent in the 4 and 5+ arms responses, where the distributions shift so that there are fewer galaxies with higher vote fractions. To correct for this bias, each of the cumulative histograms can be fit to an analytic function, and the parameters of the function modelled in terms of redshift ($z$), galaxy size ($R_{50}$) and intrinsic brightness ($M_r$). After much experimentation, a function of the following form is used to model the cumulative distributions:
\begin{equation}
f(p) = e^{kp^{c}}\mathrm{,}
\end{equation}
where $k$ and $c$ are variables fit to each of the curves. Best-fit $k$ and $c$ values are found for each of the bins, indicated by the dashed lines in Fig.~\ref{fig:function_fit}. When fitting, the cumulative histogram is sampled evenly in $\log(p)$ to avoid the fit being weighted to the steepest parts of the curves. 

After finding $k$ and $c$ for each of the bins, we attempt to quantify how these parameters change with respect to $M_r$, $\log(R_{50})$ and $z$. A $2\sigma$ clipping is applied to all of the $k$ and $c$ values to remove any fits where discrepant $k$ or $c$ values have been found. The data is then fitted using a continuous function of the following form:
\begin{equation}
\begin{split}
A_{fit}(M_r,R_{50},z) = & A_0 + A_M(f_M(-M_r)) +\\
						& A_R(f_R(\log({R_{50}}))) + A_z(f_z(z)) \mathrm{,}
\end{split}
\end{equation}
where $A$ corresponds to either $k$ or $c$ and $f_M$, $f_R$ and $f_z$ are functions that can be either logarithmic~($\log x$), linear~($x$) or exponential~($e^x$). The values $A_0$, $A_M$, $A_R$ and $A_z$ are constants that parameterise the shape of the fit with respect to each of the terms. When fitting the data, $M_r$, $\log(R_{50})$ and $z$ correspond to their respective mean values calculated using all of the galaxies in that bin. The best combination of functions is chosen by calculating $A_0$, $A_M$, $A_R$ and $A_z$ for each combination of $f_M$, $f_R$ and $f_z$, and selecting the function that has the lowest squared residuals. We then clip any values with a $>2\sigma$ residual to this fit and re-fit the data to find a final functional form for $k$ and $c$ with respect to $M_r$, $R_{50}$ and $z$. The resulting modelled cumulative histograms for the spiral arm number question are shown by the dotted lines of Fig.~\ref{fig:function_fit}. Limits are also applied to $k$ and $c$ to avoid unphysical fits at extreme values of $M_R$, $R_{50}$ and $z$,  set by the upper and lower limits of all of the fit $k$ and $c$ values within the $2\sigma$ clipping.

After finding a functional form for $k$ and $c$ with respect to $M_r$, $\log(R_{50})$ and $z$, each of the galaxies in the sample is debiased to find its equivalent value at low redshift. To do this for an individual galaxy, a cumulative histogram is estimated using $k_{\mathrm{fit}}(M_r,R_{50},z)$ and $c_{\mathrm{fit}}(M_r,R_{50},z)$, where $M_r$, $R_{50}$ and $z$ are the properties for that particular galaxy, giving the cumulative fraction for a galaxy's raw vote fraction. The equivalent cumulative histogram at $z=0.03$ (the low redshift limit of our \textit{luminosity-limited sample}) is also found, using $k_{\mathrm{fit}}(M_r,R_{50},0.03)$ and $c_{\mathrm{fit}}(M_r,R_{50},0.03)$. The vote fraction for the corresponding cumulative fraction is read off from the low redshift cumulative histogram in a similar way as in the `bin-by-bin' method, this time using the fitted curves rather than the raw histograms. This is repeated for each of the galaxies in the sample to generate a set of debiased values for the \textit{full sample} of galaxies.

As mentioned previously, function fitting avoids issues related to the discretisation of the data. However, it does introduce its own biases, as an assumption is made that the cumulative histograms can all be well-fit by a particular set of continuous functions. This may not always be the case, so we must consider which of the above methods does the best overall job of removing redshift bias. To do this, the distributions of votes for a low-redshift reference sample are compared to the distributions of higher redshift bins. Using the \textit{luminosity-limited sample}, which is free from redshift bias across all $M_r - R_{50}$ bins, a reference sample with $0.03 \leq z < 0.035$ is defined. The rest of the \textit{luminosity-limited sample} is then split into 10 redshift slices, and the total square residual of the vote fractions from both of the debiased methods are calculated with respect to the raw vote distributions of the reference sample. The method with the lowest total square residual is used to compute the final debiased values.

\subsubsection{Results from the new debiasing method}
\label{sec:debiasing_results}

\begin{figure*}
		\centering

        \includegraphics[width=0.975\textwidth]{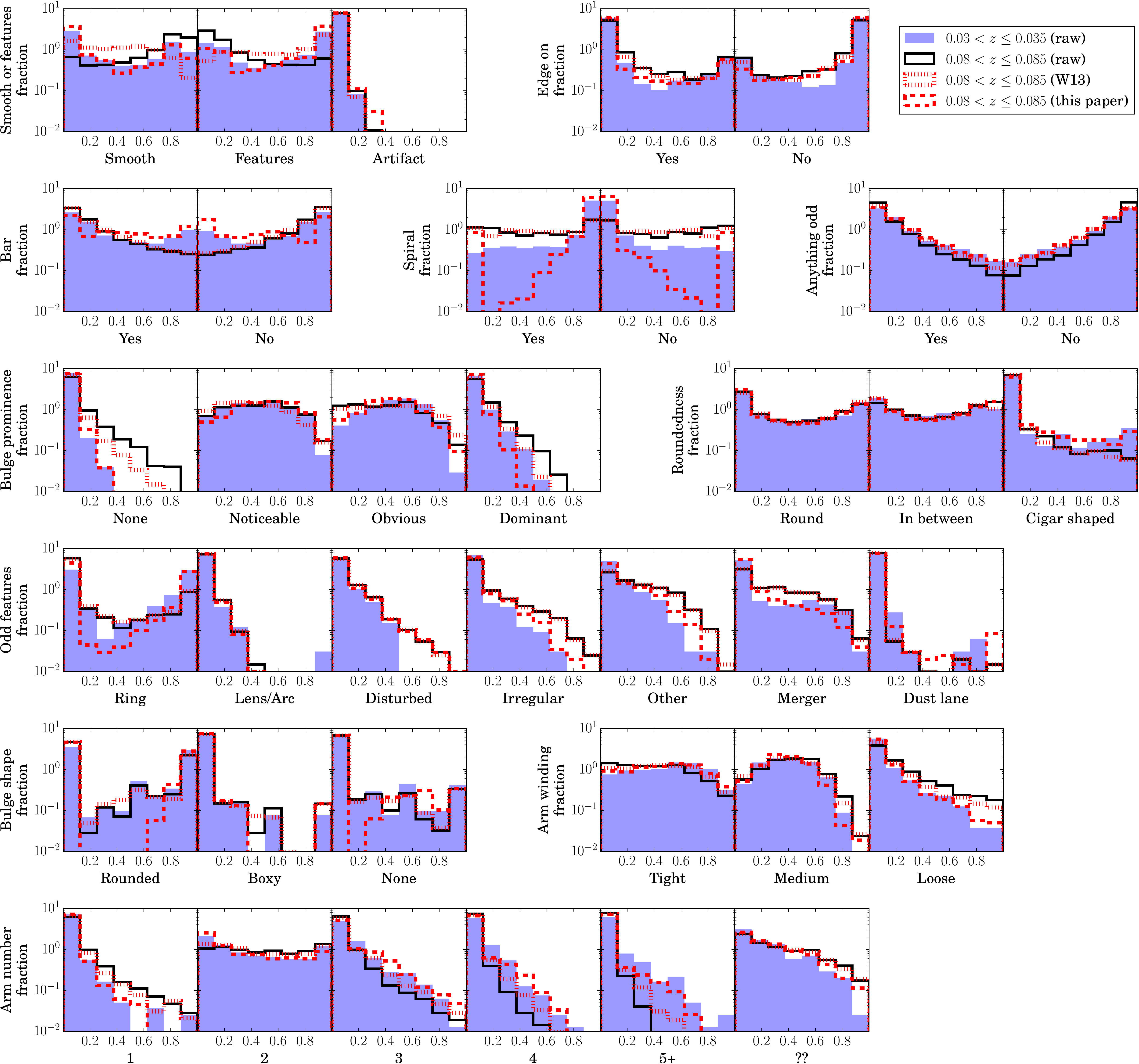}

        \caption{Vote distribution histograms for each of the answers in the GZ2 question tree. The blue filled histogram shows the distribution for galaxies with $0.03< z \leq 0.035$, which should have minimal redshift-dependent bias. The black solid, red dotted and red dashed histograms show the distribution of galaxies at $0.08 < z \leq 0.085$ using the raw, W13 debiased, and debiased data from this paper, respectively. All samples use only galaxies with $p>0.5$ (as described in Sec.~\ref{sec:sample_selection_per_question}) from the \textit{luminosity-limited sample}.}

        \label{fig:all_histograms}

\end{figure*}

As described in Sec.~\ref{sec:new_method}, the new method aims to keep the fraction of galaxies above a given threshold constant with redshift, rather than simply correcting the mean vote fractions with redshift, as shown in Fig.~\ref{fig:vote_histograms}c. To test how successful the new debiasing method is at defining populations of galaxies above a given threshold with redshift, the fraction of galaxies with $p>0.5$ for each of the questions is plotted in Fig.~\ref{fig:all_thresholds}. It can be seen that in most cases, the new debiasing method does keep the fraction of the population with $p>0.5$ constant with redshift, as expected. This effect is most evident when looking at the `spiral' question (T03 in Fig.~\ref{fig:question_tree}), in Fig.~\ref{fig:all_thresholds}d. It can be seen that the original debiasing method does not adequately remove redshift bias, with fewer galaxies exhibiting spiral structure at higher redshift. However, our new method does keep this fraction approximately constant with redshift, which means the spiral sample will be more complete if we wish to use a thresholding technique to define a sample of galaxies with spiral structure.

Fig.~\ref{fig:all_thresholds} only shows the specific example of the threshold of $p>0.5$. This does not give any insight into the overall vote fraction distribution, which can vary with redshift as shown in Fig.~\ref{fig:vote_histograms}. Therefore, overall distributions are compared for two redshift slices in Fig.~\ref{fig:all_histograms}. It can be seen that this new method does not always `match' the low and high redshift samples exactly, an effect that is most obvious in the `spiral' question. Rather than getting an excess of votes towards the middle of the distribution, an excesses are more generally seen at the tails of the distributions at $p \approx 0$ and $p \approx 1$. This is because our method preferentially matches the $p \approx 1$ end of the distribution. As can be seen by the `spiral = yes' response in Fig.~\ref{fig:all_histograms}, the top ends of the distributions are usually correctly matched; the scarcity of votes for the intermediate values of $p$ are caused by the excess of galaxies with $p=0$ that cannot be corrected. 

\subsection{Debiased data}
\label{sec:table}

The data from the new debiasing method described in this Sec.\ref{sec:new_method} is available from data.galaxyzoo.org. Alongside the raw vote fractions, our new debiased vote fractions are listed, as well as a $gz2_class$ and flags for `securely' detected spiral or elliptical galaxies (described in more detail in W13). A portion is shown in table~\ref{table:output_table} to show the form and content of the data. The table includes the weighted counts and weighted fractions from W13, with our debiased vote fractions.

\begin{table*}

\begin{tabular}{lcclcccccc}
\hline
             DR7 ID & RA          & dec         & gz2\_class   &   N\_class &   N\_votes &   wt\_count &   wt\_fraction &   debiased &   flag \\
\hline
 587732591714893851 & 11:56:10.32 & +60:31:21.1 & Sc+t        &        45 &       342 &          0 &         0     &      0     &      1 \\
 588009368545984617 & 09:00:20.26 & +52:29:39.3 & Sb+t        &        42 &       332 &          1 &         0.024 &      0.024 &      1 \\
 587732484359913515 & 12:13:29.27 & +50:44:29.4 & Ei          &        36 &       125 &         28 &         0.78  &      0.78  &      1 \\
 587741723357282317 & 12:25:00.47 & +28:33:31.0 & Sc+t        &        28 &       218 &          1 &         0.036 &      0.036 &      1 \\
 587738410866966577 & 10:44:20.73 & +14:05:04.1 & Er          &        43 &       151 &         33 &         0.767 &      0.767 &      1 \\
 587729751132209314 & 16:27:41.13 & +40:55:37.1 & Ei          &        48 &       154 &         41 &         0.861 &      0.861 &      1 \\
 587733608555216981 & 16:37:53.91 & +36:04:22.9 & Ei          &        39 &       142 &         25 &         0.649 &      0.649 &      1 \\
 587735742617616406 & 16:12:35.22 & +29:21:54.2 & Sb+t        &        35 &       282 &          0 &         0     &      0     &      1 \\
 587738574068908121 & 13:01:06.73 & +39:50:29.3 & Ei          &        50 &       158 &         42 &         0.856 &      0.856 &      1 \\
 587731870708596837 & 12:12:14.89 & +56:10:39.1 & Sb?t        &        43 &       275 &          8 &         0.194 &      0.194 &      0 \\
\hline
\end{tabular}

\caption{Example portion of the output table from the new debiasing method, showing the results from the  `smooth or features question (T11), and 'smooth answer (A0) .The full, machine-readable version of this table is available at http://data.galaxyzoo.org.}

\label{table:output_table}

\end{table*}

\section{Properties of spiral galaxies with respect to arm number}
\label{sec:results}

Despite how prevalent spiral galaxies are in the local universe, formulating a single, complete picture as to how they form and evolve is still elusive. Spiral arms are associated with enhanced levels of gas density (eg. \citealt{Grabelsky_87,EE_87b,Engargiola_03}), star-formation \citep{Seigar_02,Grosbol_12} and dust opacity \citep{Holwerda_05}. One of the key reasons why this is the case is because spiral structure can take many varied appearances. Spiral galaxies are often classified using either their Hubble type \citep{Hubble_26} or an Elmegreen-type classification scheme \citep{EE_82,EE_87}. Using the Hubble method, spiral galaxies are assigned Hubble types depending on their bulge prominences and pitch angles. More detailed classification can be applied using the de Vaucouleurs classification scheme \citep{DeV_59,DeV_63}, where the presence of more detailed structure such as diffuse, irregular spiral arms and rings can also morphologically assigned. However, the Hubble-type classification scheme and its later revisions classify spiral galaxies by their bulge prominence and their spiral arm pitch angle. These properties are weakly related \citep{Kennicutt_81,Seigar_98}: spiral arm tightness has been shown to be more strongly correlated with bulge total mass \citep{Seigar_08,Berrier_13,Davis_15}, rather than bulge-to-disk ratio. The Elmegreen-type classifications scheme instead divides galaxies into different types depending on the spiral arm structure itself, rather than any properties related to the galactic bulge. This scheme generally classifies galaxies as one of three types: grand design, multiple-armed or flocculent. Grand design spiral structure is associated with two symmetric spiral arms, whereas multiple-armed structure is associated with more than two spiral arms and flocculent galaxies have many, shorter, less well-defined arms. The distinct advantage to classifying spiral galaxies in this way is that contrasting physical mechanisms are thought to play a role in the formation of these two different types of spiral structure.

Grand design spiral structure was initially thought to be due to the presence of a density wave in a galaxy's disk \citep{Lindblad_63,Lin_64}, in which gas is `shocked' into forming stars in regions of high density in the disk. However, this mechanism is no longer favoured, as there is no evidence for the enhancement of star formation in grand design spiral galaxies compared to many-armed spiral galaxies of the same stellar mass \citep{EE_86,Dobbs_09}, or any evidence for enhancement in star formation in the individual arms of such galaxies \citep{Foyle_11,Choi_15}. Instead, it is thought that grand design spiral structure may actually occur as a result of strong bars in galaxy disks or tidal interactions \citep{Kormendy_79}. Early observational evidence supports the theory that grand design structure can be induced via interactions, with two-armed structure being favoured over many-armed structure in high density environments \citep{EE_82,EE_87,Ann_14}, and simulations showing that galaxy-galaxy interactions can lead to grand design spiral structure in galaxy disks like that seen in the local Universe \citep{Dobbs_10,Semczuk_15}. 

Unlike two-armed spiral structure, many-armed spiral structure arises readily in simulations without the requirement for a trigger from either a bar instability or a tidal interaction \citep{James_78,Sellwood_84}. Such structures require a cooling of the gas in the disk to be sustained for long periods of time \citep{Carlberg_85}. More recent simulations, taking the disk gravity into account, have shown that `flocculent' structure may actually be a transient feature of spiral galaxies, with spiral arms continually being made and destroyed \citep{Bottema_03,Grand_12b,Baba_09,Baba_13,Donghia_13}, rather than a long-lasting persistent structure.

Despite the recent advances in the simulations of these disk galaxies, the picture as to how all of the processes shape spiral galaxies still remains unclear. Grand design spiral galaxies can still reside in low density environments without the presence of bars \citep{EE_82}, meaning that they are not purely driven by these processes as described in \citet{Kormendy_79}. Additionally, the timescales of the persistence of spiral structure is still unclear, particularly as older stellar populations viewed in the infrared show very different structure to the young stellar populations viewed at optical wavelengths \citep{Block_91,Block_94,Thornley_96}. Most recent work on spiral structure have also mainly been focused on simulations of spiral structure. Putting observational constraints requires the visual inspection of the spiral arm structure in galaxy disks, so have been restricted to relatively small samples of order $\lesssim$ 2000 galaxies (eg. \citet{EE_82,EE_89,Ann_13}). We use the GZ2 vote classifications to compare the overall demographics of spiral structure in a much larger sample of SDSS galaxies, defining galaxy samples which are complete in both luminosity and stellar mass (see Sec.~\ref{sec:data} for descriptions of how these samples are defined).
\subsection{Spiral arms in Galaxy Zoo}
\label{sec:defining_the_sample}

\begin{figure}
		\centering

        \includegraphics[width=0.45\textwidth]{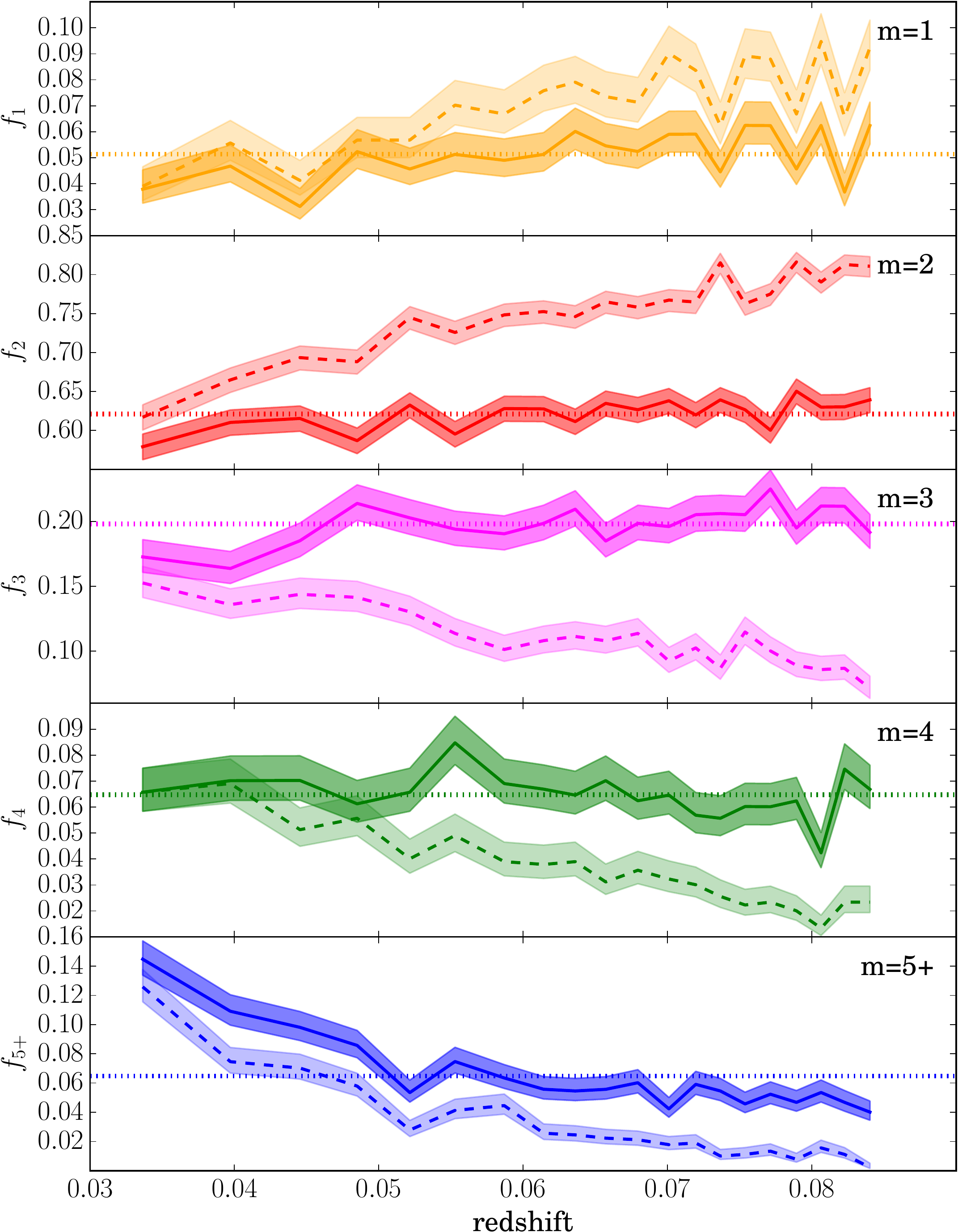}

        \caption{Fraction of galaxies in the \textit{luminosity-limited spiral sample} classified as having 1, 2, 3, 4, or 5+ spiral arms as a function of redshift. The solid lines indicates the fractions from the debiased values in this paper, and the dashed line indicates the same fractions using the raw vote fractions. Errors are calculated using the method described in \citet{Cameron_11}. The horizontal dotted lines show the mean fractions using the debiased values averaged over all of the bins.}

        \label{fig:arm_number_trend}

\end{figure}

\begin{figure*}
		\centering

        \includegraphics[width=0.9\textwidth]{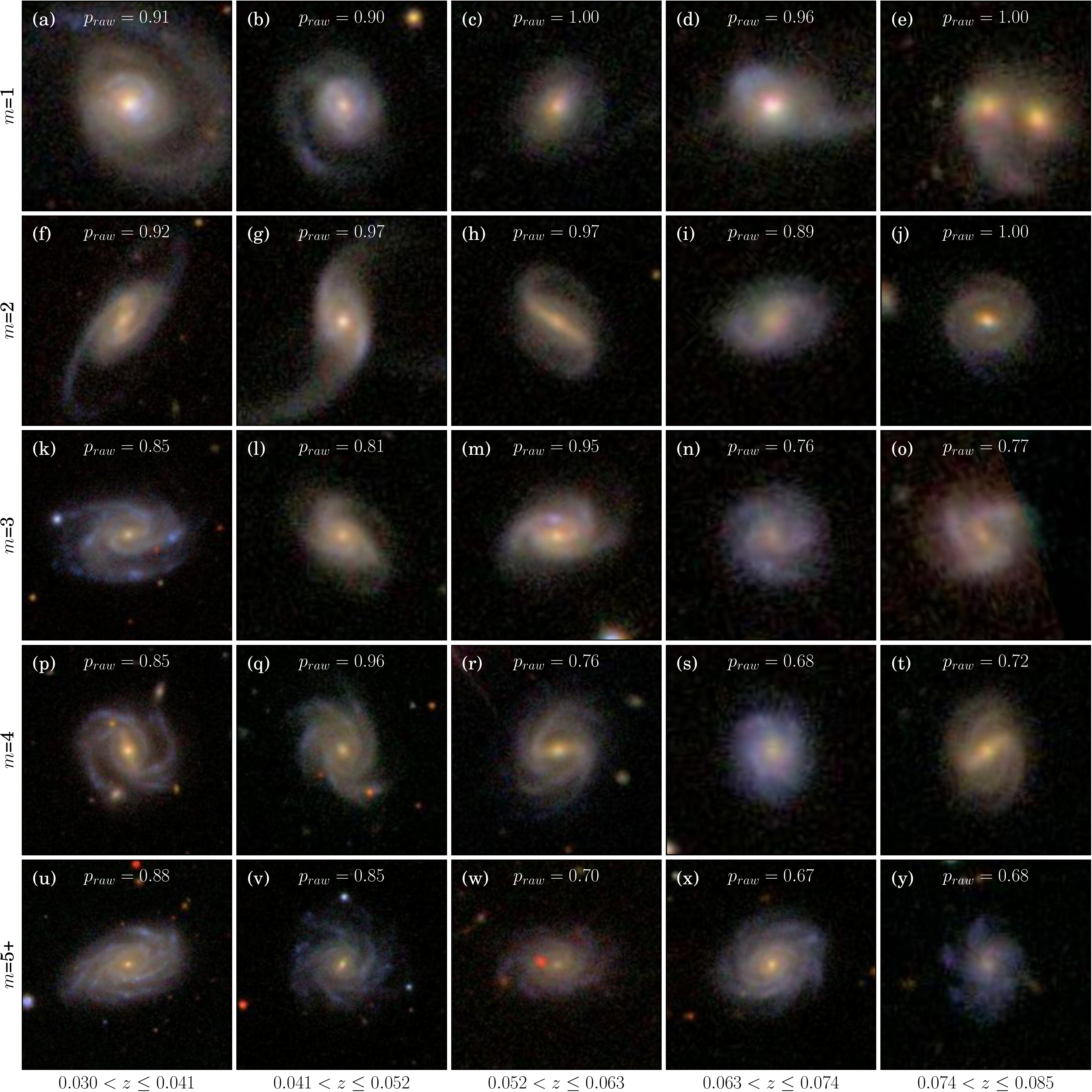}

        \caption{Galaxies classified in each of the arm number categories ($m$=1, 2, 3, 4 or 5+) for the stellar mass range $10.0 < \log{M_*/M_{\odot}} \leq 11.0$. All of the galaxies are taken from the \textit{luminosity-limited spiral sample}. Each galaxy has a debiased modal vote fraction $p_m>0.8$.}

        \label{fig:image_panel_secure}

\end{figure*}

\begin{figure*}
		\centering

        \includegraphics[width=0.9\textwidth]{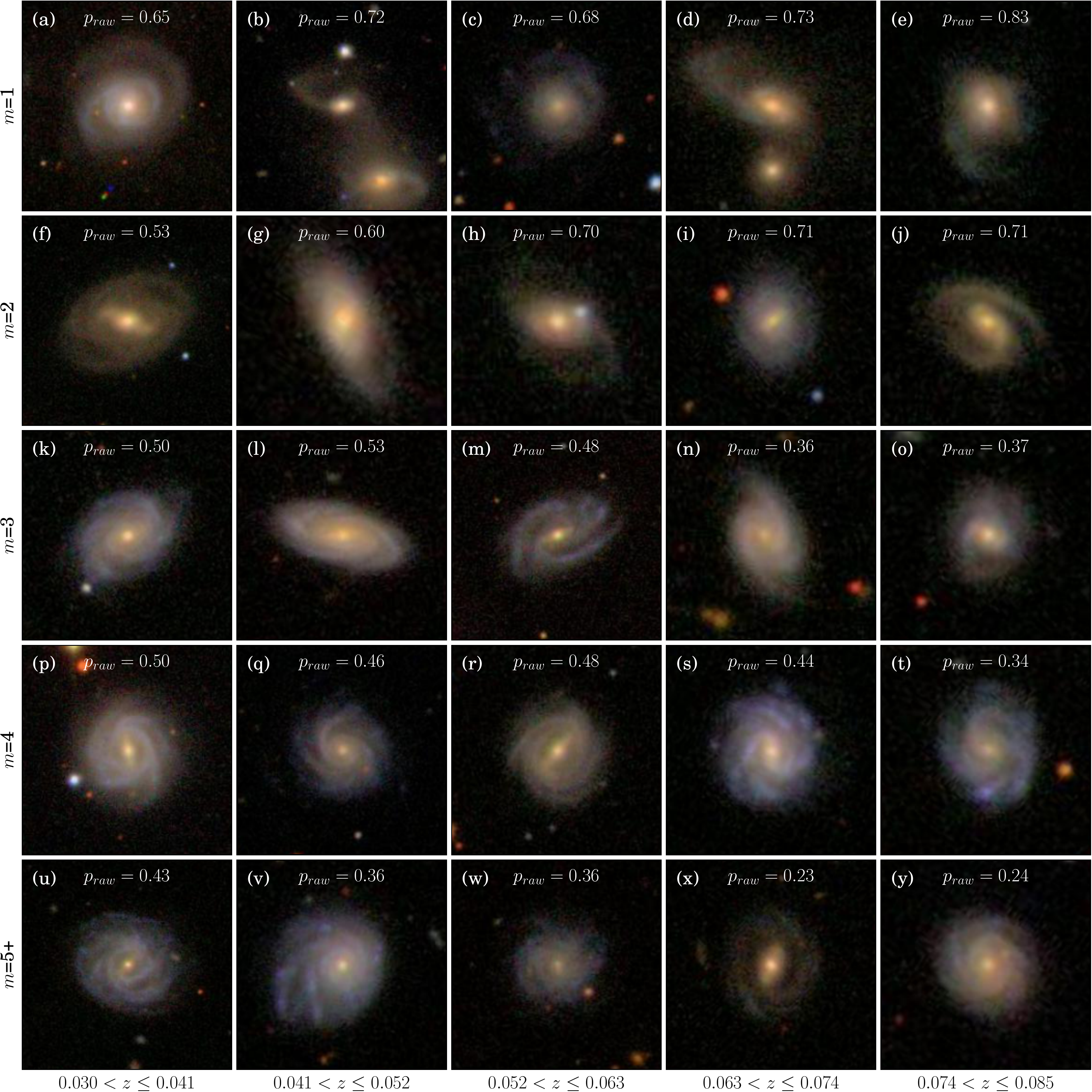}

        \caption{Galaxies classified in each of the arm number categories ($m$=1, 2, 3, 4 or 5+) for the stellar mass range $10.6 < \log{M_*/M_{\odot}} \leq 11.0$. All of the galaxies are taken from the \textit{luminosity-limited spiral sample}. Each of the galaxies is assigned to an arm number category by its modal $p_m$ value. All of the modal $p_m$-values lie in the range $0.5 < p_m \leq 0.6$.}

        \label{fig:image_panel}

\end{figure*}

In order to study how spiral properties vary, visual inspection of the number of arms in a spiral galaxy disk is required. Such classifications are provided by question T10 of the GZ2 question tree (see Fig.~\ref{fig:question_tree}). This question has six possible responses. In this case, the responses will be referred to as $m$-values, and can take the value of either 1, 2, 3, 4, 5+ or `can't tell'.

In order to compare different spiral galaxies, a secure sample of spirals must first be defined. The sample is defined by selecting galaxies with $p_{\textrm{features}} \times p_{\textrm{not edge-on}} \times p_{\textrm{spiral}} > 0.5$. A further cut is also imposed where only galaxies with $N_\mathrm{spiral} - N_\mathrm{can't \, tell} \geq 5$ are selected, meaning that \emph{at least 5 people} classified the spiral arm number of each of the spiral galaxies, reducing the effects of noise due to low numbers of classifications. The population of galaxies selected in this way from the \textit{full sample} will hereafter be referred to as the \textit{spiral sample}. The samples defined using these same cuts from the \textit{luminosity-limited sample} and \textit{stellar mass-limited sample} are referred to as the \textit{luminosity-limited spiral sample} and \textit{stellar mass-limited spiral sample}.

Each galaxy is then assigned a specific spiral arm number $m$, of either 1, 2, 3, 4 or 5+ arms, depending on which response has the highest debiased vote fraction (excluding the \textit{can't tell} response). The \emph{debiased} vote fractions for each of the arm number responses are hereafter referred to as $p_m$, where $m$ is either 1, 2, 3, 4 or 5+. Examples of some securely classified spiral galaxies are shown in Fig.~\ref{fig:image_panel_secure}, where each galaxy has a dominant vote fraction of $p_m > 0.8$. The samples of galaxies assigned to each of the different $m$-values are referred to as the \textit{arm number samples}. 

The debiasing procedure applied to this question has shifted the vote fractions for the multiple-armed ($m$=3, 4, 5+) answers upwards overall, as can be seen in Fig.~\ref{fig:arm_number_trend}. This has the effect of making each of these samples more complete with redshift, and increasing their respective overall vote fractions. However, in the $m$=5+ arms case, the sample is still somewhat incomplete, as the overall fraction of galaxies that are assigned to this category decreases with redshift. The vote fractions for $m$=5+ fall to 0 far more quickly with redshift than any of the other categories, as can be seen from the dashed line in the bottom panel of Fig.~\ref{fig:arm_number_trend}, making the  modelling of this redshift bias difficult. Despite this, the fraction of galaxies that make up the $m$=5+ category are still significantly improved compared to the sample sizes that would be defined using either the raw vote fractions or the W13 debiased vote fractions, as can be seen in from the $N$ and $f$ columns of Table \ref{table:overall_property_table}. 

\begin{table*}

\begin{tabular}{cccccccccc}
\hline
 $m$                    & $N_{\mathrm{raw}}$   & $f_{\mathrm{raw}}$   & $N_{\mathrm{W13}}$   & $f_{\mathrm{W13}}$   & $N_{\mathrm{debiased}}$   & $f_{\mathrm{debiased}}$   & $M_* (\log(M/M_{\odot}))$   & $g-i$            & $\Sigma \mathrm{(Mpc^{-2})}$  \\
 
 \hline
 Luminosity-limited   & 12554 & 1.00    & 14297 & 1.00    & 17957 & 1.00    & 10.62 (0.25) & 0.82 (0.17) & -0.24 (0.56) \\
 1                    &   563 & 0.04 &   670 & 0.05 &   926 & 0.05 & 10.63 (0.28) & 0.83 (0.19) & -0.25 (0.54) \\
 2                    &  9044 & 0.72 & 10073 & 0.7  & 11157 & 0.62 & 10.63 (0.24) & 0.86 (0.17) & -0.21 (0.57) \\
 3                    &  1778 & 0.14 &  2158 & 0.15 &  3552 & 0.2  & 10.59 (0.26) & 0.75 (0.15) & -0.28 (0.53) \\
 4                    &   615 & 0.05 &   751 & 0.05 &  1162 & 0.06 & 10.60 (0.26) & 0.74 (0.15) & -0.30 (0.51) \\
 5+                   &   554 & 0.04 &   645 & 0.05 &  1160 & 0.06 & 10.65 (0.27) & 0.75 (0.16) & -0.30 (0.53) \\
 
\hline

 Stellar mass-limited &  6683 & 1.00    &  7226 & 1.00    &  9413 & 1.00    & 10.81 (0.16) & 0.91 (0.14) & -0.18 (0.57) \\
 1                    &   290 & 0.04 &   331 & 0.05 &   500 & 0.05 & 10.84 (0.16) & 0.94 (0.14) & -0.19 (0.53) \\
 2                    &  4852 & 0.73 &  5191 & 0.72 &  6059 & 0.64 & 10.80 (0.15) & 0.94 (0.13) & -0.15 (0.59) \\
 3                    &   886 & 0.13 &   991 & 0.14 &  1654 & 0.18 & 10.82 (0.16) & 0.83 (0.12) & -0.23 (0.53) \\
 4                    &   335 & 0.05 &   366 & 0.05 &   565 & 0.06 & 10.82 (0.16) & 0.82 (0.12) & -0.25 (0.53) \\
 5+                   &   320 & 0.05 &   347 & 0.05 &   635 & 0.07 & 10.85 (0.18) & 0.82 (0.13) & -0.26 (0.53) \\
\hline

\end{tabular}

\caption{Overall properties of galaxy populations with different numbers of spiral arms. The number of galaxies with 1, 2, 3, 4 and 5+ arms are shown for both the \textit{luminosity-limited} and \textit{stellar mass-limited spiral samples}. Mean stellar masses, colours and local densities are shown for each of the populations, with $1 \sigma$ standard deviations indicated in parentheses. Errors on the mean ($\sigma / \sqrt{}N_{\mathrm{debiased}}$) are all of order $ < 0.01$.}

\label{table:overall_property_table}

\end{table*}

The main result of this debiasing is that galaxies with low vote fractions for the many-armed answers are included in the many-armed categories when they were not before. As a consequence, the population of $m$=2 galaxies is less contaminated by galaxies that actually have 3, 4 or 5+ spiral arms. This effect is illustrated in Fig.~\ref{fig:image_panel}, where a selection of spiral galaxies with $0.5 < p_m \leq 0.6$ are shown. It can be seen that the $m$=4 and $m$=5+ spiral samples at higher redshift include spiral galaxies that initially had much lower overall vote fractions. As an example, if one were to use the raw vote fractions to select `secure' galaxy samples with $p_m>0.5$, then the galaxy in Fig.~\ref{fig:image_panel}y would be unclassified, as its highest value of $p_m$ would only be 0.27 (which is actually for the $m$=4 response). Using our debiased values, it has a modal value of $p_m$=0.55 for the $m$=5+ armed response, so would be in the $m$=5+ sample. Even in the case of the less secure samples of Fig.~\ref{fig:image_panel}, the galaxies classified as $m$=4 or $m$=5+ clearly have more spiral arms than those in the $m$=2 category.

\subsection{Comparing galaxy populations}
\label{sec:comparison}

Having defined the samples of spiral galaxies in Sec.\ref{sec:defining_the_sample}, the demographics of the different galaxy populations separated by spiral arm number can be compared. For reference, mean stellar mass ($M_*$), colour ($g-i$) and local densities ($\Sigma$, as described in \citet{Baldry_06,Bamford_09}) are tabulated in the final three columns of Table~\ref{table:overall_property_table}.


\subsubsection{Comparison of sample sizes}
\label{sample_fractions}

Spiral arm multiplicity does not map exactly to a specific Elmegreen-type for two reasons. Firstly, the arm number itself does not give any indication of the prominence of spiral arms, so cannot be used to distinguish between a galaxy with many well-defined arms and one with more flocculent spiral structure, which are usually defined differently \citep{EE_82,EE_87}. The second issue is that arm structure may not necessarily be consistent at all radii \citep{Grosbol_04} or at all wavelengths \citep{Block_91,Block_94,Thornley_96} within a galaxy disk, meaning that assigning a single $m$-value of arm number may not give a complete picture of the overall spiral arm structure. The most `easy-to-map' categories may therefore be to compare the $m$=2 population with the galaxies classified as grand design, as grand design structure is usually associated with two well-defined arms across the entire disk \citep{EE_82}. In the \textit{luminosity-limited spiral sample}, $62.1 \pm 0.4 \%$ of the galaxies show two-armed spiral structure. This result is consistent with optical visual classifications \citep{EE_82} and infrared classifications \citep{Grosbol_04}, which suggest that $\sim 60\%$ of local spiral galaxies exhibit grand design spiral structure. 

\subsubsection{Stellar mass}
\label{sec:mass}

\begin{figure*}
		\centering

        \includegraphics[width=0.975\textwidth]{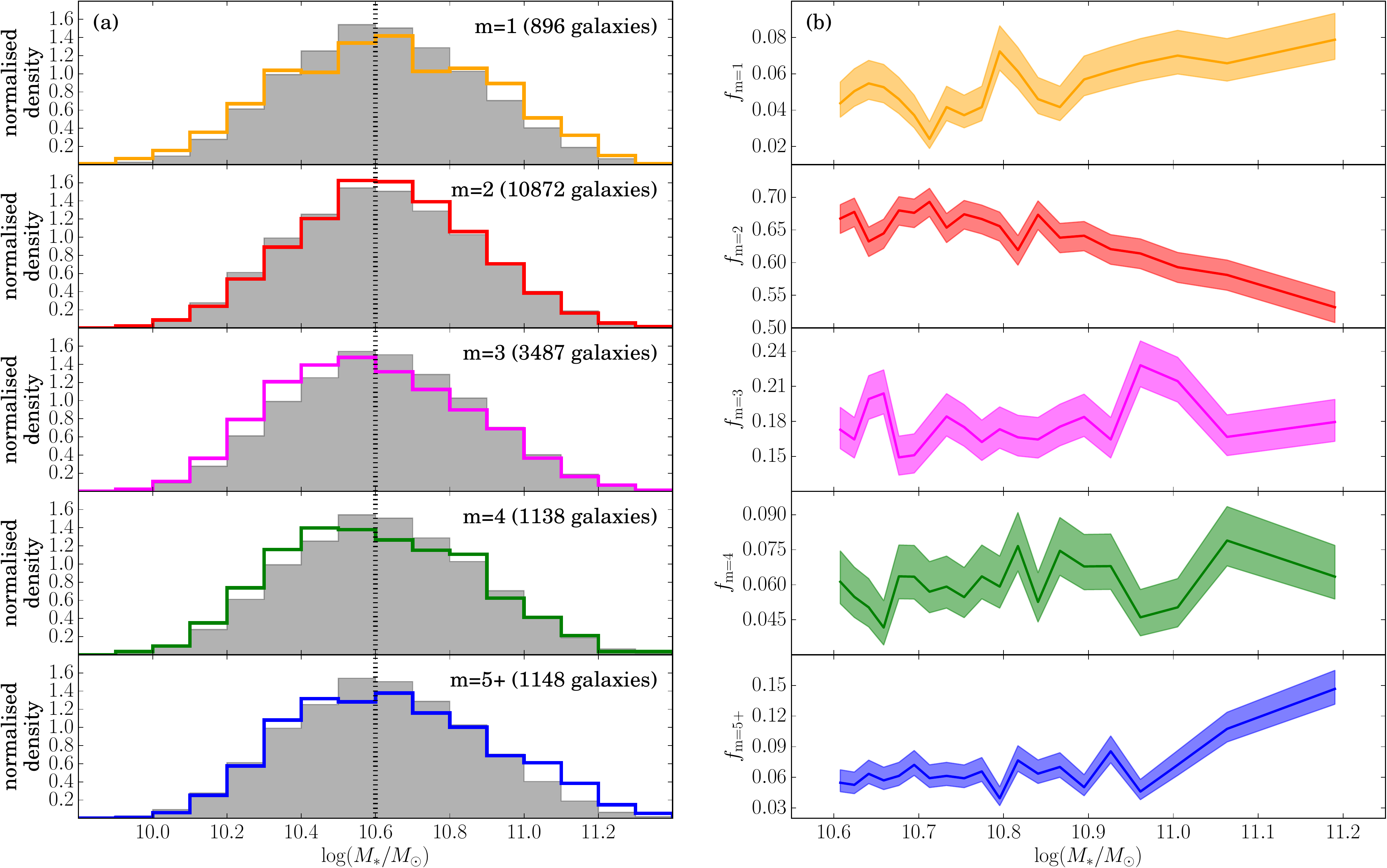}

        \caption{Left: distributions of stellar mass for the \textit{luminosity-limited spiral sample}. The solid lines indicate the distributions for each of the \textit{arm number samples} for each of arm numbers. The grey filled histograms show the equivalent distribution for all of the spiral galaxies for reference. The black dotted line indicates the stellar mass values above which the sample is complete in stellar mass. Right: fraction of the \textit{stellar mass-limited spiral sample} classified as having each spiral arm number, in 20 bins of stellar mass. The shaded regions indicate the 1$\sigma$ error calculated using the method described in \citet{Cameron_11}.}

        \label{fig:mass_plots}

\end{figure*}

Galaxy stellar mass is known to correlate with galaxy morphology \citep{Bamford_09,Kelvin_14}, and spiral galaxy Hubble type \citep{Munoz-Mateos_15}. It has been demonstrated that the \textit{central} mass of spiral galaxies can play a role in the type of spiral structure exhibited in spiral galaxies. In particular, the pitch angle of spiral arms is related to both the star-formation rate in spiral galaxies \citep{Seigar_05}, and the central mass concentration of the spiral galaxies \citep{Seigar_06,Seigar_14}. Total galaxy stellar mass has also been found to correlate with observed spiral structure, with the strength of the $m$=2 mode in spiral galaxies being stronger in galaxies with greater physical size \citep{EE_87} and stellar mass \citep{Kendall_15}. In this section we will investigate whether the total galaxy stellar mass has any influence on the number of spiral arms in spiral galaxies. 

The method for measuring stellar mass, described in \citep{Baldry_06}, uses the $u-r$ and $M_r$ values from the SDSS. To avoid contamination of galaxies with uncertain stellar masses due to poor flux detection in these bands, only galaxies with $F/\delta F > 5$ (where $F$ is the flux error in a given band, and $\delta F$ is the equivalent error on the flux) in both $u$ and $r$ are included in this analysis. The distributions of stellar mass for each of the \textit{arm number samples} are shown in Fig.~\ref{fig:mass_plots}a. The overall distributions for each of the galaxy samples show that there is little evidence for a dependence of spiral arm number with respect to host galaxy stellar mass; each of the samples contains galaxies across the entire range of stellar mass from $10.0 \lesssim \log(M_*/M_{\odot}) \lesssim 11.5$. A slight excess of low stellar mass galaxies is found in the $m$=3 and $m$=4 samples, as well as an excess of high stellar mass spiral galaxies for the $m$=5+ sample. 

The distributions of Fig.~\ref{fig:mass_plots}a show the distributions from the \textit{luminosity-limited spiral sample}, so are therefore incomplete for galaxies with lower stellar masses (see Sec.~\ref{sec:sample}) than $M_* \lesssim 10^{10.6} M_{\odot}$, indicated by the black dotted line. As we shall see in Sec.~\ref{sec:colours}, higher mass galaxies are bluer, and hence more luminous for a given stellar mass. They are thus over-represented in a at low masses in a luminosity-limited sample. To look for trends in terms of stellar mass, the overall fraction of the \textit{stellar mass-limited spiral sample} is shown in Fig.~\ref{fig:mass_plots}b. Now, it can be seen that there do appear to be some trends between spiral arm number and host galaxy stellar mass. A significant increase in the fraction of galaxies with 5+ spiral arms is observed from the overall mean value of $0.068 \pm 0.002$ to $0.15 \pm 0.02$ for the highest stellar mass bin of $\log(M_*/M_{\odot}) = 11.2 \pm 0.1$. The $m$=3 and $m$=4 samples hint at similar, but much weaker trends. Conversely, the fraction of galaxies with two spiral arms decreases from $0.642 \pm 0.004$ for the total population to $0.53 \pm 0.02$ in the highest stellar mass bin. 

One possibility why higher mass spirals may exhibit more spiral arms is that this could purely be an effect from the visual classifications. It has already been  identified that the many-armed spiral features are the most difficult to detect, so  may be more easily identifiable in the largest, brightest spiral galaxies. Spiral arms are already known to have greater amplitudes (ie. be more prominent) in galaxies with larger stellar masses \citep{Kendall_15}. It has already been demonstrated in Sec.~\ref{sec:debiasing_results} that the $m$=5+ sample is the most incomplete of the samples divided by spiral arm number. Thus, galaxies with greater stellar mass, that are therefore larger and brighter, may be preferentially put in this category, even after debiasing.

Another interesting scenario may be that the population of galaxies with the highest stellar mass are a population of unquenched spiral galaxies as in \citet{Ogle_16}. Such galaxies still have their disks intact, so have no signatures of tidal interactions. As galaxy-galaxy interactions have been linked to both the inducement of two-armed spiral structure \citep{Dobbs_10,Semczuk_15}, and the depletion of gas and therefore quenching \citep{Di_Matteo_07,Li_08}, then one may conclude that the disks of these galaxies have not been disturbed. A possible explanation for this is that lower-mass galaxies are more susceptible to environment effects \citep{Bamford_09}, so these disks are still forming stars in the transient way with multiple spiral arms, as described in Sec.~\ref{sec:results}.
\subsubsection{Local environment}
\label{sec:environment}

\begin{figure*}
		\centering

        \includegraphics[width=0.975\textwidth]{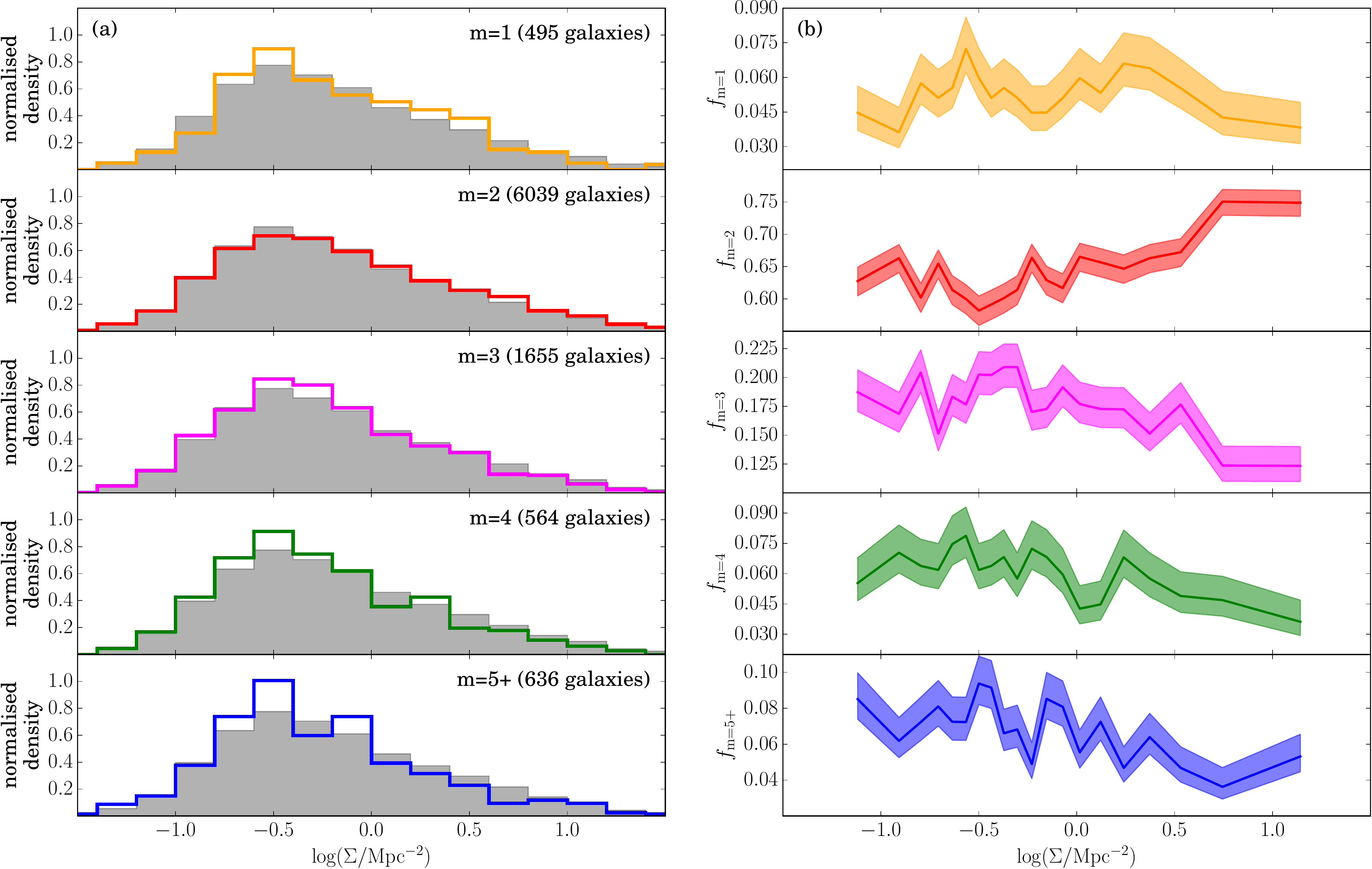}

        \caption{Left: distributions of local density ($\Sigma$) for the \textit{stellar mass-limited spiral sample}. The solid lines indicate the distributions for each of the \textit{arm number samples} for each of arm numbers. The grey filled histograms show the equivalent distribution for all of the spiral galaxies for reference. Right: fraction of the \textit{stellar mass-limited spiral sample} classified as having each spiral arm number, in 20 bins of $\Sigma$. The shaded regions indicate the 1$\sigma$ error calculated using the method described in \citet{Cameron_11}.}

        \label{fig:density_plots}

\end{figure*}

It is already well established that there is a clear dependence of the type of spiral structure that galaxies exhibit with respect to their local environment. Observational evidence from comparison of visually classified galaxies has found that grand design galaxies are more prominent in high density group environments  and in binary systems where a close companion galaxy is present \citep{EE_82,EE_87,Seigar_03,Elmegreen_11}. These results suggest that a mechanism is responsible for the transformation of spiral structure as galaxies enter the highest density environments, with a plausible explanation being that two-armed spiral structure is the result of a recent gravitational interaction. N-body modelling of galaxies has shown that two-armed spiral structure can occur as a result of galaxy-galaxy interactions \citep{Sundelius_87,Dobbs_10}. However, the timescales of the persistence of such structures are thought to be relatively short-lived \citep{Oh_08,Dobbs_10}, meaning that an enhancement in the fraction of grand design galaxies is only observed in the highest density environments where interactions can happen on a frequent enough basis to sustain such structures \citep{EE_87}.

To compare spiral arm structure as a function of environment, a mean of $\Sigma_4$ and $\Sigma_5$ is used as an estimate of local density, as in \citet{Baldry_06,Bamford_09}, denoted as $\Sigma$. $\log\Sigma$ is calculated as the mean of the density enclosed within the projected distance to the 4$^\mathrm{th}$ and 5$^\mathrm{th}$ neighbour and is hence an adaptive scale that probes both large scales outside groups and local scales within groups. 

The distributions of galaxy local densities for each of the \textit{arm number samples} are shown in Fig.~\ref{fig:density_plots}a. Here, the \textit{stellar mass-limited spiral sample} is used to define the total population, as $M_*$ and density are closely related \citep{Baldry_06}, so any biases in terms of the stellar mass distributions may have an effect on the completeness of the galaxy sample in terms of environment. The distributions show a modest dependence of spiral arm number with local density. However, as was the case for stellar mass, each of the arm number samples spans the entire range of local density defined by $\Sigma$. 

The fraction of spiral galaxies which exhibit each of the spiral arm numbers  as a function of $\log Sec.igma$ are shown in Fig.~\ref{fig:density_plots}b. A clear trend  is observed, with the number of two-armed spiral galaxies increasing for the highest values of local density from $64.3 \pm 0.5 \%$ for the overall population to $75 \pm 2 \%$ for the highest density bin of $\log Sec.igma = 1.1 \pm 0.2$. Conversely, all of the many-armed samples with $m$=3,4 or 5+ all show the opposite trends, with their respective fractions decreasing with $\Sigma$. These results therefore seem to be in qualitative agreement with \citet{EE_82} and \citet{Ann_14}, in which the fraction of galaxies displaying grand design spiral structure increases in the highest density environments. As the increase seems to be most distinct in the very highest densities, this could be indicative that two-armed spiral structure is a short-lived phase induced by galaxy-galaxy interactions, as described in \citet{EE_83}. Interestingly, there is no clear enhancement in the fraction of galaxies with a single spiral arm at the highest densities, as found in \citet{Casteels_13}. However, \citet{Casteels_13} found the most significant enhancements in $m$=1 galaxies when a galaxies have a close companion, which is not probed by our measure of environment. A more complte analysis of spiral structure with local environment, accounting for both interaction probabilities and local density will need to be considered to look for more significant trends of spiral arm structure with environment. With our large, clean samples of galaxies with measurements of arm number, we plan to take a more thorough analysis of of spiral structure with environment in a future paper.

\subsubsection{Galaxy colours}
\label{sec:colours}

\begin{figure*}
		\centering

        \includegraphics[width=0.975\textwidth]{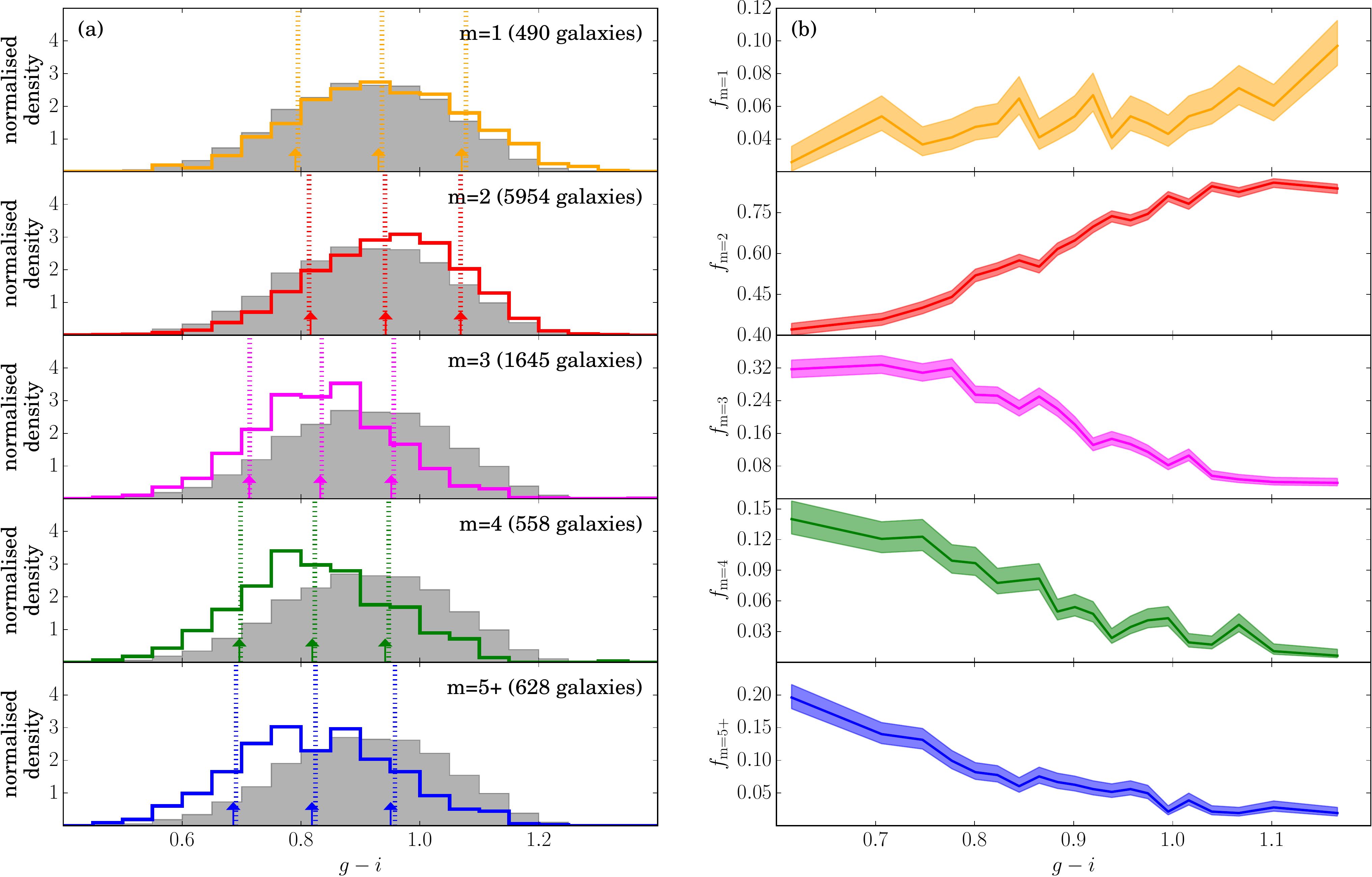}

        \caption{Left: distributions of $g-i$ colour for the \textit{stellar mass-limited spiral sample}. The solid lines indicate the distributions for each of the \textit{arm number samples} for each of arm numbers. The grey filled histograms show the equivalent distribution for all of the spiral galaxies for reference. Right: fraction of the \textit{stellar mass-limited spiral sample} classified as having each spiral arm number, in 20 bins of $g-i$. The shaded regions indicate the 1$\sigma$ error calculated using the method described in \citet{Cameron_11}. The arrows indicate the mean and 1 standard deviation scatter for samples matched in stellar mass.}

        \label{fig:colour_plots}

\end{figure*}

Colours primarily indicate stellar population ages in galaxies, although dust extinction can also have an effect. Star-formation properties have been hypothesised to correlate with spiral arm properties, where galaxies with more prominent spiral arms show enhanced star-formation \citep{Seigar_02,Kendall_15}. The presence of a density wave in a galaxy disk has been proposed as a method of inducing star formation, but the lack of evidence for a clear enhancement of star formation in grand design spiral galaxies \citep{EE_86,Foyle_10,Willett_15} or a clear age gradient within spiral arms \citep{Foyle_11,Dobbs_14,Choi_15} suggests that this is not the case. 

Galaxy colour is already known to relate to stellar mass (eg. \citealt{Kauffmann_04,Baldry_06}), environment (eg. \citealt{Kauffmann_04,Baldry_04} and overall galaxy morphology (eg. \citealt{Aaronson_78,Glass_84,Bamford_09}). As spiral arms are associated with recent star formation, and also the presence of dust \citep{Grosbol_12}, we expect their properties to correlate with colour. Thus, galaxy colour correlates with the presence of spiral arms, with spiral galaxies being bluer in colour than ellipticals \citep{Bamford_09,Schawinski_14}. The colour distributions are now compared to look for any trends with recent star formation history in Fig.~\ref{fig:colour_plots}a. The colours that are plotted here are the SDSS $g-i$ optical colours, which should probe recent star formation in galaxies. To avoid contamination from poor detections, only the galaxies with $F/\delta F > 5$ in both $g$ and $i$ are included. Unlike the distributions of local density and stellar mass, a strong trend is found between colour and arm multiplicity. The two-armed spiral galaxies show the reddest overall colours, with mean $g-i$ of 0.94 and a standard deviation of 0.13 in the \textit{stellar mass-limited spiral sample}. The $m$=3, 4 and 5+ armed samples have corresponding colours of 0.83, 0.82 and 0.82, with corresponding standard deviations of 0.12, 0.12 and 0.13. Thus, each of the many-armed spiral samples is $\approx$1 standard deviation bluer than the two armed spiral galaxy population. A population of barred red spirals in Galaxy Zoo have been found before in \citep{Masters_10b}. As grand design spiral spiral structure is associated with two spiral arms \citep{EE_82}, this red spiral galaxy population may be composed of strongly barred, grand design spiral galaxies.

To further compare the overall galaxy colours, the fraction of the \textit{stellar mass-limited spiral sample} with each of the spiral arm numbers with respect to $g-i$ is shown in Fig.~\ref{fig:colour_plots}b. Here, a clear trend is observed with the fraction of galaxies displaying two spiral arms with respect to colour. In the bluest bin ($g-i=0.67 \pm 0.07$), only $32 \pm 2 \%$ of galaxies have two spiral arms; in the reddest bin $g-i=1.17 \pm 0.05$), $84\pm2\%$ have two spiral arms. 
\begin{figure*}
		\centering

        \includegraphics[width=0.975\textwidth]{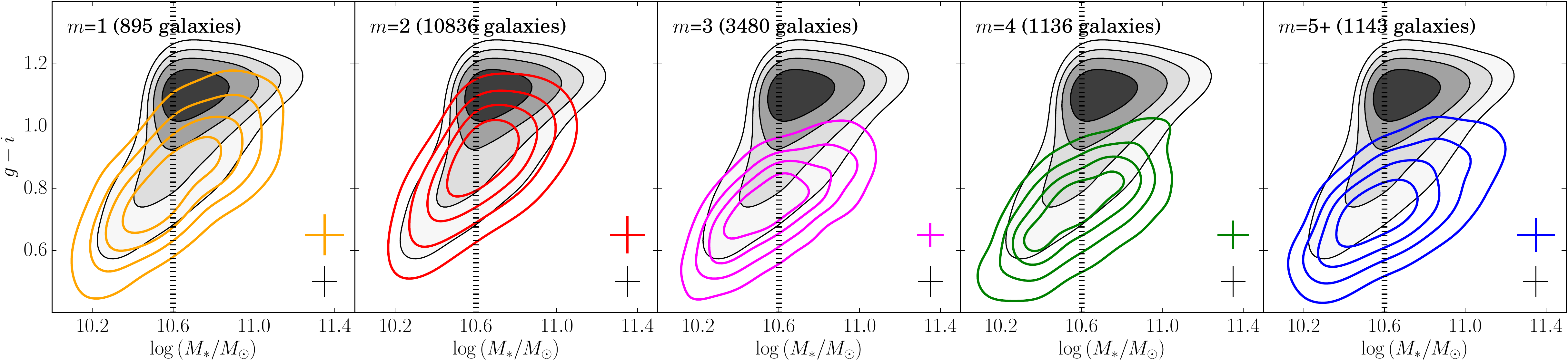}

        \caption{Stellar mass vs. $g-i$ colour galaxy samples classified by spiral arm number in the \textit{luminosity-limited sample}. The black dotted line indicates where the sample is incomplete in terms of stellar mass. The greyscale shaded contours show the total \textit{stellar mass-limited sample} for all morphologies, whereas the solid lines show the distributions for each \textit{arm number sample}. The contours are plotted with a kernel density estimate, of bandwidth optimised using 5-fold cross validation, and the selected bandwidths are displayed in the bottom-right corner of each plot. The contour levels show the regions enclosing 20\%, 40\%, 60\%, and 80\% of the points.}

        \label{fig:mass-colour}

\end{figure*}

As described above, a strong dependence of colour with stellar mass is well-known (eg. \citealt{Baldry_06}). However, as described in \ref{sec:mass}, our samples only show very weak trends with stellar mass. To test whether any of the colour differences between the samples can be attributed to differences in stellar mass, $g-i$ colour is plotted against stellar mass in Fig.~\ref{fig:mass-colour}. The results show that the colour differences cannot be explained by the stellar mass differences between our \textit{arm number samples}: for a given stellar mass, the many-armed spiral galaxies are much bluer in the $g-i$ band. The samples were also matched in terms of stellar mass, and the mean and standard deviations are indicated by the arrows in Fig.~\ref{fig:colour_plots}. The colour differences are still $\approx 1$ standard deviation bluer in the many-armed spirals compared to the two-armed spirals, after matching the samples by stellar mass.

\begin{figure*}
		\centering

        \includegraphics[width=0.975\textwidth]{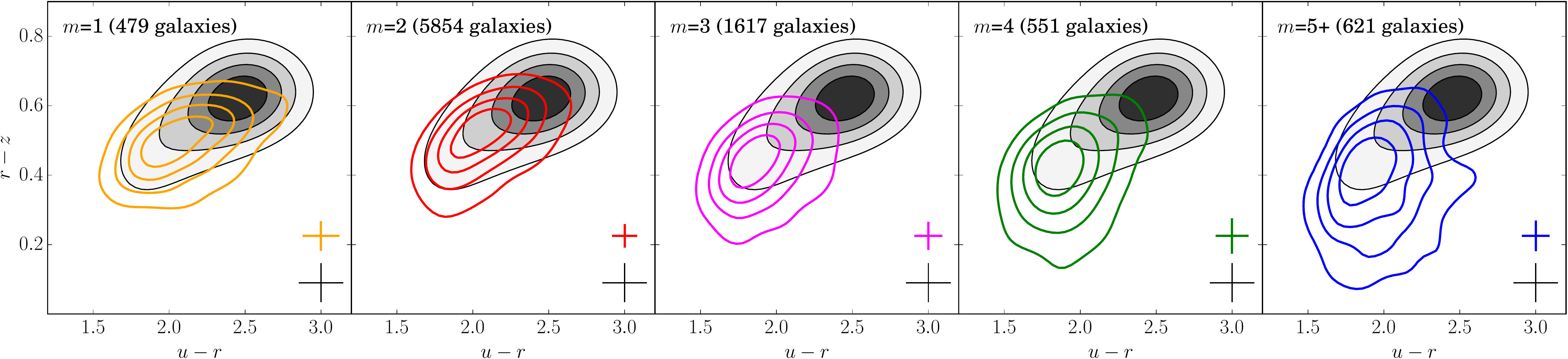}

        \caption{$u-r$ vs. $r-z$ colours for each of the \textit{arm number samples} taken from the \textit{stellar mass limited spiral sample}. The greyscale shaded contours show the entire \textit{stellar mass-limited sample}, irrespective of morphology, whereas the solid lines indicate the same distribution for each \textit{arm number sample}. Contours are plotted using a kernel density estimate,with bandwidths optimised using 5-fold cross validation. The selected bandwidths are displayed in the bottom-right corner of each plot. The contour levels show the regions enclosing 20\%, 40\%, 60\%, and 80\% of the data for each sample.}

        \label{fig:colour-colour}

\end{figure*}

\begin{figure}
		\centering

        \includegraphics[width=0.45\textwidth]{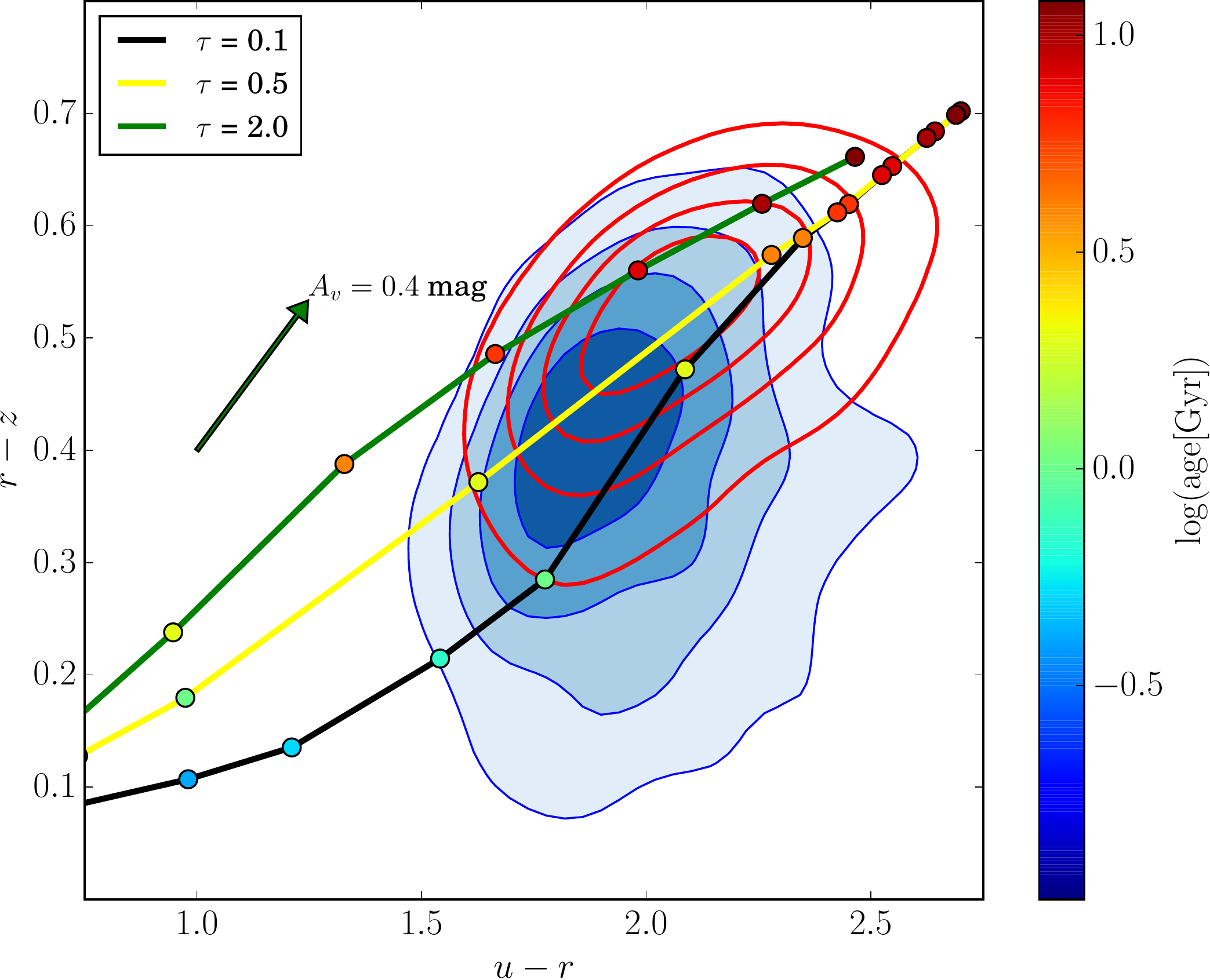}

        \caption{Contour plots for the $m$=2 (red solid contours)and $m$=5+ (blue filled contours) samples as in Fig.~\ref{fig:colour-colour}. Three evolutionary tracks for \citet{BC_03} stellar population models with different quenching timescales ($\tau$) are plotted in black, yellow and green lines, indicated in the plot legend. Each point is coloured by the relative age of the SFH models ($t$), indicated in the colourbar. The green arrow indicates how the evolutionary curves would change colors with dust extinction $A_{V}$.}

        \label{fig:cc_w_sfh}

\end{figure}

Using a single colour only gives a broad indication as to how the star formation properties of galaxies differ. To try to gain a more detailed understanding of the star-formation in each of the \textit{arm number samples} the $u-r$ and $r-z$ bands are compared for each of the different arm numbers, and the results are plotted in Fig.~\ref{fig:colour-colour}. Similar cuts in $F/ \delta F$ to the $u$, $r$ and $z$ bands as described in \ref{sec:mass} are used to define the samples. It can be seen that the differences are stronger in $r-z$ than in $u-r$. The most significant differences are observed between the $m=2$ and $m=5+$ samples, where there is a significant offset in $r-z$ for a given $u-r$. 

In order to gain an insight into how star-formation can have affected the galaxy colours, the $m$=2 and $m$=5+ $u-r$ vs. $r-z$ distributions are plotted are plotted  in Fig.~\ref{fig:cc_w_sfh}, with $\tau$-model SFHs for reference from \citet{BC_03} (see Sec.~\ref{sec:SEDs} for details). The SFH models are for a quenching galaxy, defined with two parameters, $t$ and $\tau$, where $t$ is the time of quenching onset and $\tau$ is the quenching timescale (a shorter $\tau$ means a faster quenching). For each of the three timescales, the dust extinction $A_{V}$ is set to 0. The plot indicates that both populations are consistent with SFH model colours, but that the quenching process is much longer in the $m$=2 population (indicated by a longer $\tau$) than in the $m$=5+ population. The $m$=5+ population has therefore undergone a shorter, more recent phase of star formation. We also see a significant population of galaxies that are red in $u-r$ and blue in $r-z$, which cannot be explained by a $\tau$-model, even with a quickly declining SFR. A model with a recent, short burst superimposed on a longer, smoother SFH may be more suitable.

The role of dust must also be considered. A reference dust attenuation of $A_v$=0.4 is shown by the green arrow of Fig.~\ref{fig:cc_w_sfh}. The arrow indicates that extinction by dust could account for the some of differences in the colours of the galaxies if the attenuation is higher in the $m$=2 population than the $m$=5+ population. However, such a scenario would seem unlikely, as dust opacity is greater within spiral arms \citep{Holwerda_05}. Therefore, one would expect that the spiral galaxies with more spiral arms to have a greater level of dust attenuation overall. Galaxies with greater levels of dust attenuation are also expected to have lower SFR \citep{Garn_10}, with the most passive spiral galaxies being the most dust deficient \citep{Rowlands_12}. It is therefore unlikely that dust attenuation in spiral galaxies could play a significant role unless the SFRs of two-armed spiral galaxies are significantly enhanced, which is not found to be the case \citep{Willett_15}.

Recent simulations of disks in spiral galaxies have proposed that flocculent spiral structure can be sustained for long periods of time, of order $>10$~Gyr, \citep{Fujii_11,Donghia_13}, with spiral arms being frequently made and broken. Our results suggest instead that flocculent spiral structure is a short-lived phase, associated with a recent star formation event. Simulations frequently model disks in isolation, so may not account for all processes, eg. the effects that environment can have on the inducement or transformation of the spiral structure in local galaxies.

To gain a more complete understanding of the effects of dust and star-formation with respect to spiral arm number, further SFH models will be explored in a later paper. SFH models with more than a single component will be considered, as well as how the presence of bars and gas content affect the SFHs of the different galaxies. 
\section{Conclusions}
\label{sec:conclusions}

In this paper, the demographics of local spiral galaxies have been compared with respect to spiral arm number, in order to gain an understanding of any significant differences in the physical processes responsible for their spiral structure. We make use of visual classifications of SDSS galaxies from GZ2. In order to obtain complete and clean samples, we have developed a new method to account for redshift-dependent bias. This corrects the vote fractions to ensure sample completeness, and avoid contamination between separate classes of galaxies. The method will also be applicable to further studies of Galaxy Zoo data, and potentially other citizen science projects.

A new debiasing method has been developed to remove the effects of redshift-dependent classification bias in Galaxy Zoo data. The method was required for the multiple-answer questions in Galaxy Zoo, where the previously defined debiasing method did not effectively remove redshift bias, leading to sample contamination from incorrectly classified galaxies. In this paper, we studied the arm-number question, which is a multiple-answer question, where the rarer many-armed samples were incomplete, and the two-armed category suffered from sample contamination. The new method  was successful in making the samples more complete with redshift in this case.

Using the resulting classifications, the distributions of environment, stellar mass and colour were compared for spiral galaxies with different numbers of arms. We found that the most massive galaxies favour many-armed spiral structure, which may be indicative that their disks have not have been sufficiently perturbed to induce two-armed spiral structure. An enhancement in the fraction of two-armed spiral galaxies was observed in the highest density environments, indicating that galaxy-galaxy interactions could play a role in the inducement of two-armed spiral structure. By comparing optical colours, we find that two-armed galaxies are much redder in colour than galaxies with many spiral arms. Although many-armed spiral galaxies display similar $u-r$ colours, the $r-z$ colours are distinctly redder in the two-armed galaxy population. These colours are indicative of a recent, rapidly quenched ($\lesssim$0.1 Gyr) burst of star-formation, suggesting that many-armed spiral structure is a short-lived phase in galaxy disks, whereas star-formation in two-armed spiral structure persists over much longer timescales.

\section{Acknowledgments}

The data in this paper are the result of the efforts of the Galaxy Zoo 2 volunteers, without whom none of this work would be possible. Their efforts are individually acknowledged at \url{http://authors.galaxyzoo.org}.

RH acknowledges a studentship from the Science and Technology Funding Council.

The development of Galaxy Zoo was supported in part by the Alfred P. Sloan foundation and the Leverhulme Trust.

Cross validation and kernel density methods made use of \texttt{scikit-learn} \citep{scikit-learn} and \texttt{astroML} \citep{astroML}. This publication made extensive use of the \texttt{scipy} Python module \citep{scipy}, including the \texttt{scipy.optimize} module for curve fitting and minimisation. This research also made use of \texttt{Astropy}, a community-developed core Python package for Astronomy \citep{astropy}.

Funding for the Sloan Digital Sky Survey (SDSS) and SDSS-II has been provided by the Alfred P. Sloan Foundation, the Participating Institutions, the National Science Foundation, the U.S.Department of Energy, the National Aeronautics and Space Ad
ministration, the Japanese Monbukagakusho, and the Max Planck Society, and the Higher Education Funding Council for England. The SDSS Web site is http://www.sdss.org/. The SDSS is managed by the Astrophysical Research Consortium (ARC) for the Participating Institutions. The Participating Institutions are the American Museum of Natural History, Astrophysical Institute Potsdam, University of Basel, University of Cambridge, Case Western Reserve University, The University of Chicago, Drexel University, Fermilab, the Institute for Advanced Study, the Japan Participation Group, The Johns Hopkins University, the Joint Institute for Nuclear Astrophysics, the Kavli Institute for Particle Astrophysics and Cosmology, the Korean Scientist Group, the Chinese Academy of Sciences (LAMOST), Los Alamos National Laboratory, the Max-Planck-Institute for Astronomy (MPIA), the Max-Planck-Institute for Astrophysics (MPA), New Mexico State University, Ohio State University, University of Pittsburgh, University of Portsmouth, Princeton University, the United States Naval Observatory, and the University of Washington.

\bibliographystyle{mn2e}
\bibliography{jun15}

\begin{thebibliography}{109}
\expandafter\ifx\csname natexlab\endcsname\relax\def\natexlab#1{#1}\fi

\bibitem[{{Aaronson}(1978)}]{Aaronson_78}
{Aaronson} M., 1978, ApJL, 221, L103

\bibitem[{{Abazajian} {et~al}\mbox{.}(2009){Abazajian}, {Adelman-McCarthy},
  {Ag{\"u}eros}, {Allam}, {Allende Prieto}, {An}, {Anderson}, {Anderson},
  {Annis}, {Bahcall}, \& et~al.}]{Abazijian_09}
{Abazajian} K.~N. {et~al.}, 2009, ApJS, 182, 543

\bibitem[{{Ann}(2014)}]{Ann_14}
{Ann} H.~B., 2014, Journal of Korean Astronomical Society, 47, 1

\bibitem[{{Ann} \& {Lee}(2013)}]{Ann_13}
{Ann} H.~B., {Lee} H.-R., 2013, Journal of Korean Astronomical Society, 46, 141

\bibitem[{{Astropy Collaboration} {et~al}\mbox{.}(2013){Astropy Collaboration},
  {Robitaille}, {Tollerud}, {Greenfield}, {Droettboom}, {Bray}, {Aldcroft},
  {Davis}, {Ginsburg}, {Price-Whelan}, {Kerzendorf}, {Conley}, {Crighton},
  {Barbary}, {Muna}, {Ferguson}, {Grollier}, {Parikh}, {Nair}, {Unther},
  {Deil}, {Woillez}, {Conseil}, {Kramer}, {Turner}, {Singer}, {Fox}, {Weaver},
  {Zabalza}, {Edwards}, {Azalee Bostroem}, {Burke}, {Casey}, {Crawford},
  {Dencheva}, {Ely}, {Jenness}, {Labrie}, {Lim}, {Pierfederici}, {Pontzen},
  {Ptak}, {Refsdal}, {Servillat}, \& {Streicher}}]{astropy}
{Astropy Collaboration} {et~al.}, 2013, AAP, 558, A33

\bibitem[{{Baba} {et~al}\mbox{.}(2009){Baba}, {Asaki}, {Makino}, {Miyoshi},
  {Saitoh}, \& {Wada}}]{Baba_09}
{Baba} J., {Asaki} Y., {Makino} J., {Miyoshi} M., {Saitoh} T.~R., {Wada} K.,
  2009, ApJ, 706, 471

\bibitem[{{Baba}, {Saitoh} \& {Wada}(2013){Baba}, {Saitoh}, \&
  {Wada}}]{Baba_13}
{Baba} J., {Saitoh} T.~R., {Wada} K., 2013, ApJ, 763, 46

\bibitem[{{Baldry} {et~al}\mbox{.}(2004){Baldry}, {Balogh}, {Bower},
  {Glazebrook}, \& {Nichol}}]{Baldry_04}
{Baldry} I.~K., {Balogh} M.~L., {Bower} R., {Glazebrook} K., {Nichol} R.~C.,
  2004, in American Institute of Physics Conference Series, Vol. 743, The New
  Cosmology: Conference on Strings and Cosmology, {Allen} R.~E., {Nanopoulos}
  D.~V., {Pope} C.~N., eds., pp. 106--119

\bibitem[{{Baldry} {et~al}\mbox{.}(2006){Baldry}, {Balogh}, {Bower},
  {Glazebrook}, {Nichol}, {Bamford}, \& {Budavari}}]{Baldry_06}
{Baldry} I.~K., {Balogh} M.~L., {Bower} R.~G., {Glazebrook} K., {Nichol} R.~C.,
  {Bamford} S.~P., {Budavari} T., 2006, MNRAS, 373, 469

\bibitem[{{Bamford} {et~al}\mbox{.}(2009){Bamford}, {Nichol}, {Baldry}, {Land},
  {Lintott}, {Schawinski}, {Slosar}, {Szalay}, {Thomas}, {Torki}, {Andreescu},
  {Edmondson}, {Miller}, {Murray}, {Raddick}, \& {Vandenberg}}]{Bamford_09}
{Bamford} S.~P. {et~al.}, 2009, MNRAS, 393, 1324

\bibitem[{{Berrier} {et~al}\mbox{.}(2013){Berrier}, {Davis}, {Kennefick},
  {Kennefick}, {Seigar}, {Barrows}, {Hartley}, {Shields}, {Bentz}, \&
  {Lacy}}]{Berrier_13}
{Berrier} J.~C. {et~al.}, 2013, ApJ, 769, 132

\bibitem[{{Blanton} \& {Roweis}(2007)}]{Blanton_07}
{Blanton} M.~R., {Roweis} S., 2007, AJ, 133, 734

\bibitem[{{Block} {et~al}\mbox{.}(1994){Block}, {Bertin}, {Stockton},
  {Grosbol}, {Moorwood}, \& {Peletier}}]{Block_94}
{Block} D.~L., {Bertin} G., {Stockton} A., {Grosbol} P., {Moorwood} A.~F.~M.,
  {Peletier} R.~F., 1994, A\&A, 288

\bibitem[{{Block} \& {Wainscoat}(1991)}]{Block_91}
{Block} D.~L., {Wainscoat} R.~J., 1991, Nature, 353, 48

\bibitem[{{Bottema}(2003)}]{Bottema_03}
{Bottema} R., 2003, MNRAS, 344, 358

\bibitem[{{Bruzual} \& {Charlot}(2003)}]{BC_03}
{Bruzual} G., {Charlot} S., 2003, MNRAS, 344, 1000

\bibitem[{{Calzetti} {et~al}\mbox{.}(2000){Calzetti}, {Armus}, {Bohlin},
  {Kinney}, {Koornneef}, \& {Storchi-Bergmann}}]{Calzetti_00}
{Calzetti} D., {Armus} L., {Bohlin} R.~C., {Kinney} A.~L., {Koornneef} J.,
  {Storchi-Bergmann} T., 2000, ApJ, 533, 682

\bibitem[{{Cameron}(2011)}]{Cameron_11}
{Cameron} E., 2011, PASA, 28, 128

\bibitem[{{Cappellari} \& {Copin}(2003)}]{Cappellari_03}
{Cappellari} M., {Copin} Y., 2003, MNRAS, 342, 345

\bibitem[{{Carlberg} \& {Freedman}(1985)}]{Carlberg_85}
{Carlberg} R.~G., {Freedman} W.~L., 1985, ApJ, 298, 486

\bibitem[{{Casteels} {et~al}\mbox{.}(2013){Casteels}, {Bamford}, {Skibba},
  {Masters}, {Lintott}, {Keel}, {Schawinski}, {Nichol}, \&
  {Smith}}]{Casteels_13}
{Casteels} K.~R.~V. {et~al.}, 2013, MNRAS, 429, 1051

\bibitem[{{Chabrier}(2003)}]{Chabrier_03}
{Chabrier} G., 2003, PASP, 115, 763

\bibitem[{{Cheung} {et~al}\mbox{.}(2013){Cheung}, {Athanassoula}, {Masters},
  {Nichol}, {Bosma}, {Bell}, {Faber}, {Koo}, {Lintott}, {Melvin}, {Schawinski},
  {Skibba}, \& {Willett}}]{Cheung_13}
{Cheung} E. {et~al.}, 2013, ApJ, 779, 162

\bibitem[{{Choi} {et~al}\mbox{.}(2015){Choi}, {Dalcanton}, {Williams}, {Weisz},
  {Skillman}, {Fouesneau}, \& {Dolphin}}]{Choi_15}
{Choi} Y., {Dalcanton} J.~J., {Williams} B.~F., {Weisz} D.~R., {Skillman}
  E.~D., {Fouesneau} M., {Dolphin} A.~E., 2015, ApJ, 810, 9

\bibitem[{{Darg} {et~al}\mbox{.}(2010{\natexlab{a}}){Darg}, {Kaviraj},
  {Lintott}, {Schawinski}, {Sarzi}, {Bamford}, {Silk}, {Andreescu}, {Murray},
  {Nichol}, {Raddick}, {Slosar}, {Szalay}, {Thomas}, \&
  {Vandenberg}}]{Darg_10b}
{Darg} D.~W. {et~al.}, 2010{\natexlab{a}}, MNRAS, 401, 1552

\bibitem[{{Darg} {et~al}\mbox{.}(2010{\natexlab{b}}){Darg}, {Kaviraj},
  {Lintott}, {Schawinski}, {Sarzi}, {Bamford}, {Silk}, {Proctor}, {Andreescu},
  {Murray}, {Nichol}, {Raddick}, {Slosar}, {Szalay}, {Thomas}, \&
  {Vandenberg}}]{Darg_10a}
{Darg} D.~W. {et~al.}, 2010{\natexlab{b}}, MNRAS, 401, 1043

\bibitem[{{Davis} {et~al}\mbox{.}(2015){Davis}, {Kennefick}, {Kennefick},
  {Westfall}, {Shields}, {Flatman}, {Hartley}, {Berrier}, {Martinsson}, \&
  {Swaters}}]{Davis_15}
{Davis} B.~L. {et~al.}, 2015, ApJL, 802, L13

\bibitem[{{Davis} \& {Hayes}(2014)}]{Davis_14}
{Davis} D.~R., {Hayes} W.~B., 2014, ApJ, 790, 87

\bibitem[{{de Vaucouleurs}(1959)}]{DeV_59}
{de Vaucouleurs} G., 1959, Handbuch der Physik, 53, 275

\bibitem[{{de Vaucouleurs}(1963)}]{DeV_63}
{de Vaucouleurs} G., 1963, ApJS, 8, 31

\bibitem[{{Di Matteo} {et~al}\mbox{.}(2007){Di Matteo}, {Combes}, {Melchior},
  \& {Semelin}}]{Di_Matteo_07}
{Di Matteo} P., {Combes} F., {Melchior} A.-L., {Semelin} B., 2007, A\&A, 468,
  61

\bibitem[{{Dieleman}, {Willett} \& {Dambre}(2015){Dieleman}, {Willett}, \&
  {Dambre}}]{Dieleman_15}
{Dieleman} S., {Willett} K.~W., {Dambre} J., 2015, MNRAS, 450, 1441

\bibitem[{{Dobbs} \& {Baba}(2014)}]{Dobbs_14}
{Dobbs} C., {Baba} J., 2014, PASA, 31, e035

\bibitem[{{Dobbs} \& {Pringle}(2009)}]{Dobbs_09}
{Dobbs} C.~L., {Pringle} J.~E., 2009, MNRAS, 396, 1579

\bibitem[{{Dobbs} {et~al}\mbox{.}(2010){Dobbs}, {Theis}, {Pringle}, \&
  {Bate}}]{Dobbs_10}
{Dobbs} C.~L., {Theis} C., {Pringle} J.~E., {Bate} M.~R., 2010, MNRAS, 403, 625

\bibitem[{{Doi} {et~al}\mbox{.}(2010){Doi}, {Tanaka}, {Fukugita}, {Gunn},
  {Yasuda}, {Ivezi{\'c}}, {Brinkmann}, {de Haars}, {Kleinman}, {Krzesinski}, \&
  {French Leger}}]{Doi_10}
{Doi} M. {et~al.}, 2010, AJ, 139, 1628

\bibitem[{{D'Onghia}, {Vogelsberger} \& {Hernquist}(2013){D'Onghia},
  {Vogelsberger}, \& {Hernquist}}]{Donghia_13}
{D'Onghia} E., {Vogelsberger} M., {Hernquist} L., 2013, ApJ, 766, 34

\bibitem[{{Duncan} {et~al}\mbox{.}(2014){Duncan}, {Conselice}, {Mortlock},
  {Hartley}, {Guo}, {Ferguson}, {Dav{\'e}}, {Lu}, {Ownsworth}, {Ashby},
  {Dekel}, {Dickinson}, {Faber}, {Giavalisco}, {Grogin}, {Kocevski},
  {Koekemoer}, {Somerville}, \& {White}}]{Duncan_14}
{Duncan} K. {et~al.}, 2014, MNRAS, 444, 2960

\bibitem[{{Elmegreen} \& {Elmegreen}(1983)}]{EE_83}
{Elmegreen} B.~G., {Elmegreen} D.~M., 1983, ApJ, 267, 31

\bibitem[{{Elmegreen} \& {Elmegreen}(1986)}]{EE_86}
{Elmegreen} B.~G., {Elmegreen} D.~M., 1986, ApJ, 311, 554

\bibitem[{{Elmegreen} \& {Elmegreen}(1987{\natexlab{a}})}]{EE_87b}
{Elmegreen} B.~G., {Elmegreen} D.~M., 1987{\natexlab{a}}, ApJ, 320, 182

\bibitem[{{Elmegreen} \& {Elmegreen}(1989)}]{EE_89}
{Elmegreen} B.~G., {Elmegreen} D.~M., 1989, ApJ, 342, 677

\bibitem[{{Elmegreen} \& {Elmegreen}(1982)}]{EE_82}
{Elmegreen} D.~M., {Elmegreen} B.~G., 1982, MNRAS, 201, 1021

\bibitem[{{Elmegreen} \& {Elmegreen}(1987{\natexlab{b}})}]{EE_87}
{Elmegreen} D.~M., {Elmegreen} B.~G., 1987{\natexlab{b}}, ApJ, 314, 3

\bibitem[{{Elmegreen} {et~al}\mbox{.}(2011){Elmegreen}, {Elmegreen}, {Yau},
  {Athanassoula}, {Bosma}, {Buta}, {Helou}, {Ho}, {Gadotti}, {Knapen},
  {Laurikainen}, {Madore}, {Masters}, {Meidt}, {Men{\'e}ndez-Delmestre},
  {Regan}, {Salo}, {Sheth}, {Zaritsky}, {Aravena}, {Skibba}, {Hinz}, {Laine},
  {Gil de Paz}, {Mu{\~n}oz-Mateos}, {Seibert}, {Mizusawa}, {Kim}, \& {Erroz
  Ferrer}}]{Elmegreen_11}
{Elmegreen} D.~M. {et~al.}, 2011, ApJ, 737, 32

\bibitem[{{Engargiola} {et~al}\mbox{.}(2003){Engargiola}, {Plambeck},
  {Rosolowsky}, \& {Blitz}}]{Engargiola_03}
{Engargiola} G., {Plambeck} R.~L., {Rosolowsky} E., {Blitz} L., 2003, ApJS,
  149, 343

\bibitem[{{Foyle} {et~al}\mbox{.}(2011){Foyle}, {Rix}, {Dobbs}, {Leroy}, \&
  {Walter}}]{Foyle_11}
{Foyle} K., {Rix} H.-W., {Dobbs} C.~L., {Leroy} A.~K., {Walter} F., 2011, ApJ,
  735, 101

\bibitem[{{Foyle} {et~al}\mbox{.}(2010){Foyle}, {Rix}, {Walter}, \&
  {Leroy}}]{Foyle_10}
{Foyle} K., {Rix} H.-W., {Walter} F., {Leroy} A.~K., 2010, ApJ, 725, 534

\bibitem[{{Fujii} {et~al}\mbox{.}(2011){Fujii}, {Baba}, {Saitoh}, {Makino},
  {Kokubo}, \& {Wada}}]{Fujii_11}
{Fujii} M.~S., {Baba} J., {Saitoh} T.~R., {Makino} J., {Kokubo} E., {Wada} K.,
  2011, ApJ, 730, 109

\bibitem[{{Galloway} {et~al}\mbox{.}(2015){Galloway}, {Willett}, {Fortson},
  {Cardamone}, {Schawinski}, {Cheung}, {Lintott}, {Masters}, {Melvin}, \&
  {Simmons}}]{Galloway_15}
{Galloway} M.~A. {et~al.}, 2015, MNRAS, 448, 3442

\bibitem[{{Garn} \& {Best}(2010)}]{Garn_10}
{Garn} T., {Best} P.~N., 2010, MNRAS, 409, 421

\bibitem[{{Glass}(1984)}]{Glass_84}
{Glass} I.~S., 1984, MNRAS, 211, 461

\bibitem[{{Grabelsky} {et~al}\mbox{.}(1987){Grabelsky}, {Cohen}, {Bronfman},
  {Thaddeus}, \& {May}}]{Grabelsky_87}
{Grabelsky} D.~A., {Cohen} R.~S., {Bronfman} L., {Thaddeus} P., {May} J., 1987,
  ApJ, 315, 122

\bibitem[{{Grand}, {Kawata} \& {Cropper}(2012){Grand}, {Kawata}, \&
  {Cropper}}]{Grand_12b}
{Grand} R.~J.~J., {Kawata} D., {Cropper} M., 2012, MNRAS, 426, 167

\bibitem[{{Grosb{\o}l} \& {Dottori}(2012)}]{Grosbol_12}
{Grosb{\o}l} P., {Dottori} H., 2012, AAP, 542, A39

\bibitem[{{Grosb{\o}l}, {Patsis} \& {Pompei}(2004){Grosb{\o}l}, {Patsis}, \&
  {Pompei}}]{Grosbol_04}
{Grosb{\o}l} P., {Patsis} P.~A., {Pompei} E., 2004, A\&A, 423, 849

\bibitem[{{Holwerda} {et~al}\mbox{.}(2005){Holwerda}, {Gonz{\'a}lez}, {van der
  Kruit}, \& {Allen}}]{Holwerda_05}
{Holwerda} B.~W., {Gonz{\'a}lez} R.~A., {van der Kruit} P.~C., {Allen} R.~J.,
  2005, AAP, 444, 109

\bibitem[{{Hubble}(1926)}]{Hubble_26}
{Hubble} E.~P., 1926, ApJ, 64

\bibitem[{{Huertas-Company} {et~al}\mbox{.}(2011){Huertas-Company}, {Aguerri},
  {Bernardi}, {Mei}, \& {S{\'a}nchez Almeida}}]{Huertas_11}
{Huertas-Company} M., {Aguerri} J.~A.~L., {Bernardi} M., {Mei} S., {S{\'a}nchez
  Almeida} J., 2011, AAP, 525, A157

\bibitem[{{James} \& {Sellwood}(1978)}]{James_78}
{James} R.~A., {Sellwood} J.~A., 1978, MNRAS, 182, 331

\bibitem[{Jones {et~al}\mbox{.}(2001--)Jones, Oliphant, Peterson,
  {et~al.}}]{scipy}
Jones E., Oliphant T., Peterson P., {et~al.}, 2001--, {SciPy}: Open source
  scientific tools for {Python}. [Online; accessed 2016-03-03]

\bibitem[{{Kauffmann} {et~al}\mbox{.}(2004){Kauffmann}, {White}, {Heckman},
  {M{\'e}nard}, {Brinchmann}, {Charlot}, {Tremonti}, \&
  {Brinkmann}}]{Kauffmann_04}
{Kauffmann} G., {White} S.~D.~M., {Heckman} T.~M., {M{\'e}nard} B.,
  {Brinchmann} J., {Charlot} S., {Tremonti} C., {Brinkmann} J., 2004, MNRAS,
  353, 713

\bibitem[{{Kelvin} {et~al}\mbox{.}(2014{\natexlab{a}}){Kelvin}, {Driver},
  {Robotham}, {Graham}, {Phillipps}, {Agius}, {Alpaslan}, {Baldry}, {Bamford},
  {Bland-Hawthorn}, {Brough}, {Brown}, {Colless}, {Conselice}, {Hopkins},
  {Liske}, {Loveday}, {Norberg}, {Pimbblet}, {Popescu}, {Prescott}, {Taylor},
  \& {Tuffs}}]{Kelvin_14b}
{Kelvin} L.~S. {et~al.}, 2014{\natexlab{a}}, MNRAS, 439, 1245

\bibitem[{{Kelvin} {et~al}\mbox{.}(2014{\natexlab{b}}){Kelvin}, {Driver},
  {Robotham}, {Taylor}, {Graham}, {Alpaslan}, {Baldry}, {Bamford}, {Bauer},
  {Bland-Hawthorn}, {Brown}, {Colless}, {Conselice}, {Holwerda}, {Hopkins},
  {Lara-L{\'o}pez}, {Liske}, {L{\'o}pez-S{\'a}nchez}, {Loveday}, {Norberg},
  {Phillipps}, {Popescu}, {Prescott}, {Sansom}, \& {Tuffs}}]{Kelvin_14}
{Kelvin} L.~S. {et~al.}, 2014{\natexlab{b}}, MNRAS, 444, 1647

\bibitem[{{Kendall}, {Clarke} \& {Kennicutt}(2015){Kendall}, {Clarke}, \&
  {Kennicutt}}]{Kendall_15}
{Kendall} S., {Clarke} C., {Kennicutt} R.~C., 2015, MNRAS, 446, 4155

\bibitem[{{Kennicutt}(1981)}]{Kennicutt_81}
{Kennicutt}, Jr. R.~C., 1981, AJ, 86, 1847

\bibitem[{{Kennicutt}(1989)}]{Kennicutt_89}
{Kennicutt}, Jr. R.~C., 1989, ApJ, 344, 685

\bibitem[{{Kormendy} \& {Norman}(1979)}]{Kormendy_79}
{Kormendy} J., {Norman} C.~A., 1979, ApJ, 233, 539

\bibitem[{{Land} {et~al}\mbox{.}(2008){Land}, {Slosar}, {Lintott}, {Andreescu},
  {Bamford}, {Murray}, {Nichol}, {Raddick}, {Schawinski}, {Szalay}, {Thomas},
  \& {Vandenberg}}]{Land_08}
{Land} K. {et~al.}, 2008, MNRAS, 388, 1686

\bibitem[{{Li} {et~al}\mbox{.}(2008){Li}, {Kauffmann}, {Heckman}, {Jing}, \&
  {White}}]{Li_08}
{Li} C., {Kauffmann} G., {Heckman} T.~M., {Jing} Y.~P., {White} S.~D.~M., 2008,
  MNRAS, 385, 1903

\bibitem[{{Lin} \& {Shu}(1964)}]{Lin_64}
{Lin} C.~C., {Shu} F.~H., 1964, ApJ, 140, 646

\bibitem[{{Lindblad}(1963)}]{Lindblad_63}
{Lindblad} B., 1963, Stockholms Observatoriums Annaler, 22

\bibitem[{{Lintott} {et~al}\mbox{.}(2011){Lintott}, {Schawinski}, {Bamford},
  {Slosar}, {Land}, {Thomas}, {Edmondson}, {Masters}, {Nichol}, {Raddick},
  {Szalay}, {Andreescu}, {Murray}, \& {Vandenberg}}]{Lintott_11}
{Lintott} C. {et~al.}, 2011, MNRAS, 410, 166

\bibitem[{{Lintott} {et~al}\mbox{.}(2008){Lintott}, {Schawinski}, {Slosar},
  {Land}, {Bamford}, {Thomas}, {Raddick}, {Nichol}, {Szalay}, {Andreescu},
  {Murray}, \& {Vandenberg}}]{Lintott_08}
{Lintott} C.~J. {et~al.}, 2008, MNRAS, 389, 1179

\bibitem[{{Maller}(2008)}]{Maller_08}
{Maller} A.~H., 2008, in Astronomical Society of the Pacific Conference Series,
  Vol. 396, Formation and Evolution of Galaxy Disks, {Funes} J.~G., {Corsini}
  E.~M., eds., p. 251

\bibitem[{{Masters} {et~al}\mbox{.}(2010{\natexlab{a}}){Masters}, {Mosleh},
  {Romer}, {Nichol}, {Bamford}, {Schawinski}, {Lintott}, {Andreescu},
  {Campbell}, {Crowcroft}, {Doyle}, {Edmondson}, {Murray}, {Raddick}, {Slosar},
  {Szalay}, \& {Vandenberg}}]{Masters_10b}
{Masters} K.~L. {et~al.}, 2010{\natexlab{a}}, MNRAS, 405, 783

\bibitem[{{Masters} {et~al}\mbox{.}(2010{\natexlab{b}}){Masters}, {Nichol},
  {Bamford}, {Mosleh}, {Lintott}, {Andreescu}, {Edmondson}, {Keel}, {Murray},
  {Raddick}, {Schawinski}, {Slosar}, {Szalay}, {Thomas}, \&
  {Vandenberg}}]{Masters_10a}
{Masters} K.~L. {et~al.}, 2010{\natexlab{b}}, MNRAS, 404, 792

\bibitem[{{Masters} {et~al}\mbox{.}(2012){Masters}, {Nichol}, {Haynes}, {Keel},
  {Lintott}, {Simmons}, {Skibba}, {Bamford}, {Giovanelli}, \&
  {Schawinski}}]{Masters_12}
{Masters} K.~L. {et~al.}, 2012, MNRAS, 424, 2180

\bibitem[{{Masters} {et~al}\mbox{.}(2011){Masters}, {Nichol}, {Hoyle},
  {Lintott}, {Bamford}, {Edmondson}, {Fortson}, {Keel}, {Schawinski}, {Smith},
  \& {Thomas}}]{Masters_11}
{Masters} K.~L. {et~al.}, 2011, MNRAS, 411, 2026

\bibitem[{{Melvin} {et~al}\mbox{.}(2014){Melvin}, {Masters}, {Lintott},
  {Nichol}, {Simmons}, {Bamford}, {Casteels}, {Cheung}, {Edmondson}, {Fortson},
  {Schawinski}, {Skibba}, {Smith}, \& {Willett}}]{Melvin_14}
{Melvin} T. {et~al.}, 2014, MNRAS, 438, 2882

\bibitem[{{Mu{\~n}oz-Mateos} {et~al}\mbox{.}(2015){Mu{\~n}oz-Mateos}, {Sheth},
  {Regan}, {Kim}, {Laine}, {Erroz-Ferrer}, {Gil de Paz}, {Comeron}, {Hinz},
  {Laurikainen}, {Salo}, {Athanassoula}, {Bosma}, {Bouquin}, {Schinnerer},
  {Ho}, {Zaritsky}, {Gadotti}, {Madore}, {Holwerda}, {Men{\'e}ndez-Delmestre},
  {Knapen}, {Meidt}, {Querejeta}, {Mizusawa}, {Seibert}, {Laine}, \&
  {Courtois}}]{Munoz-Mateos_15}
{Mu{\~n}oz-Mateos} J.~C. {et~al.}, 2015, ApJS, 219, 3

\bibitem[{{Nair} \& {Abraham}(2010)}]{Nair_10}
{Nair} P.~B., {Abraham} R.~G., 2010, ApJS, 186, 427

\bibitem[{{Ogle} {et~al}\mbox{.}(2016){Ogle}, {Lanz}, {Nader}, \&
  {Helou}}]{Ogle_16}
{Ogle} P.~M., {Lanz} L., {Nader} C., {Helou} G., 2016, ApJ, 817, 109

\bibitem[{{Oh} {et~al}\mbox{.}(2008){Oh}, {Kim}, {Lee}, \& {Kim}}]{Oh_08}
{Oh} S.~H., {Kim} W.-T., {Lee} H.~M., {Kim} J., 2008, ApJ, 683, 94

\bibitem[{Pedregosa {et~al}\mbox{.}(2011)Pedregosa, Varoquaux, Gramfort,
  Michel, Thirion, Grisel, Blondel, Prettenhofer, Weiss, Dubourg, Vanderplas,
  Passos, Cournapeau, Brucher, Perrot, \& Duchesnay}]{scikit-learn}
Pedregosa F. {et~al.}, 2011, Journal of Machine Learning Research, 12, 2825

\bibitem[{{Peng}, {Maiolino} \& {Cochrane}(2015){Peng}, {Maiolino}, \&
  {Cochrane}}]{Peng_15}
{Peng} Y., {Maiolino} R., {Cochrane} R., 2015, Nature, 521, 192

\bibitem[{{Rowlands} {et~al}\mbox{.}(2012){Rowlands}, {Dunne}, {Maddox},
  {Bourne}, {Gomez}, {Kaviraj}, {Bamford}, {Brough}, {Charlot}, {da Cunha},
  {Driver}, {Eales}, {Hopkins}, {Kelvin}, {Nichol}, {Sansom}, {Sharp}, {Smith},
  {Temi}, {van der Werf}, {Baes}, {Cava}, {Cooray}, {Croom}, {Dariush}, {de
  Zotti}, {Dye}, {Fritz}, {Hopwood}, {Ibar}, {Ivison}, {Liske}, {Loveday},
  {Madore}, {Norberg}, {Popescu}, {Rigby}, {Robotham}, {Rodighiero}, {Seibert},
  \& {Tuffs}}]{Rowlands_12}
{Rowlands} K. {et~al.}, 2012, MNRAS, 419, 2545

\bibitem[{{Schawinski} {et~al}\mbox{.}(2014){Schawinski}, {Urry}, {Simmons},
  {Fortson}, {Kaviraj}, {Keel}, {Lintott}, {Masters}, {Nichol}, {Sarzi},
  {Skibba}, {Treister}, {Willett}, {Wong}, \& {Yi}}]{Schawinski_14}
{Schawinski} K. {et~al.}, 2014, MNRAS, 440, 889

\bibitem[{{Schmidt}(1959)}]{Schmidt_59}
{Schmidt} M., 1959, ApJ, 129, 243

\bibitem[{{Seigar}(2005)}]{Seigar_05}
{Seigar} M.~S., 2005, MNRAS, 361, L20

\bibitem[{{Seigar} {et~al}\mbox{.}(2006){Seigar}, {Bullock}, {Barth}, \&
  {Ho}}]{Seigar_06}
{Seigar} M.~S., {Bullock} J.~S., {Barth} A.~J., {Ho} L.~C., 2006, ApJ, 645,
  1012

\bibitem[{{Seigar}, {Chorney} \& {James}(2003){Seigar}, {Chorney}, \&
  {James}}]{Seigar_03}
{Seigar} M.~S., {Chorney} N.~E., {James} P.~A., 2003, MNRAS, 342, 1

\bibitem[{{Seigar} {et~al}\mbox{.}(2014){Seigar}, {Davis}, {Berrier}, \&
  {Kennefick}}]{Seigar_14}
{Seigar} M.~S., {Davis} B.~L., {Berrier} J., {Kennefick} D., 2014, ApJ, 795, 90

\bibitem[{{Seigar} \& {James}(1998)}]{Seigar_98}
{Seigar} M.~S., {James} P.~A., 1998, MNRAS, 299, 685

\bibitem[{{Seigar} \& {James}(2002)}]{Seigar_02}
{Seigar} M.~S., {James} P.~A., 2002, MNRAS, 337, 1113

\bibitem[{{Seigar} {et~al}\mbox{.}(2008){Seigar}, {Kennefick}, {Kennefick}, \&
  {Lacy}}]{Seigar_08}
{Seigar} M.~S., {Kennefick} D., {Kennefick} J., {Lacy} C.~H.~S., 2008, ApJL,
  678, L93

\bibitem[{{Sellwood} \& {Carlberg}(1984)}]{Sellwood_84}
{Sellwood} J.~A., {Carlberg} R.~G., 1984, ApJ, 282, 61

\bibitem[{{Semczuk} \& {Lokas}(2015)}]{Semczuk_15}
{Semczuk} M., {Lokas} E.~L., 2015, ArXiv e-prints

\bibitem[{{Simmons} {et~al}\mbox{.}(2014){Simmons}, {Melvin}, {Lintott},
  {Masters}, {Willett}, {Keel}, {Smethurst}, {Cheung}, {Nichol}, {Schawinski},
  {Rutkowski}, {Kartaltepe}, {Bell}, {Casteels}, {Conselice}, {Almaini},
  {Ferguson}, {Fortson}, {Hartley}, {Kocevski}, {Koekemoer}, {McIntosh},
  {Mortlock}, {Newman}, {Ownsworth}, {Bamford}, {Dahlen}, {Faber},
  {Finkelstein}, {Fontana}, {Galametz}, {Grogin}, {Gr{\"u}tzbauch}, {Guo},
  {H{\"a}u{\ss}ler}, {Jek}, {Kaviraj}, {Lucas}, {Peth}, {Salvato}, {Wiklind},
  \& {Wuyts}}]{Simmons_14}
{Simmons} B.~D. {et~al.}, 2014, MNRAS, 445, 3466

\bibitem[{{Skibba} {et~al}\mbox{.}(2009){Skibba}, {Bamford}, {Nichol},
  {Lintott}, {Andreescu}, {Edmondson}, {Murray}, {Raddick}, {Schawinski},
  {Slosar}, {Szalay}, {Thomas}, \& {Vandenberg}}]{Skibba_09}
{Skibba} R.~A. {et~al.}, 2009, MNRAS, 399, 966

\bibitem[{{Smethurst} {et~al}\mbox{.}(2015){Smethurst}, {Lintott}, {Simmons},
  {Schawinski}, {Marshall}, {Bamford}, {Fortson}, {Kaviraj}, {Masters},
  {Melvin}, {Nichol}, {Skibba}, \& {Willett}}]{Smethurst_15}
{Smethurst} R.~J. {et~al.}, 2015, MNRAS, 450, 435

\bibitem[{{Strauss} {et~al}\mbox{.}(2002){Strauss}, {Weinberg}, {Lupton},
  {Narayanan}, {Annis}, {Bernardi}, {Blanton}, {Burles}, {Connolly},
  {Dalcanton}, {Doi}, {Eisenstein}, {Frieman}, {Fukugita}, {Gunn},
  {Ivezi{\'c}}, {Kent}, {Kim}, {Knapp}, {Kron}, {Munn}, {Newberg}, {Nichol},
  {Okamura}, {Quinn}, {Richmond}, {Schlegel}, {Shimasaku}, {SubbaRao},
  {Szalay}, {Vanden Berk}, {Vogeley}, {Yanny}, {Yasuda}, {York}, \&
  {Zehavi}}]{Strauss_02}
{Strauss} M.~A. {et~al.}, 2002, AJ, 124, 1810

\bibitem[{{Sundelius} {et~al}\mbox{.}(1987){Sundelius}, {Thomasson},
  {Valtonen}, \& {Byrd}}]{Sundelius_87}
{Sundelius} B., {Thomasson} M., {Valtonen} M.~J., {Byrd} G.~G., 1987, A\&A,
  174, 67

\bibitem[{{Thornley}(1996)}]{Thornley_96}
{Thornley} M.~D., 1996, ApJL, 469, L45

\bibitem[{{Tojeiro} {et~al}\mbox{.}(2013){Tojeiro}, {Masters}, {Richards},
  {Percival}, {Bamford}, {Maraston}, {Nichol}, {Skibba}, \&
  {Thomas}}]{Tojeiro_13}
{Tojeiro} R. {et~al.}, 2013, MNRAS, 432, 359

\bibitem[{{Toomre}(1981)}]{Toomre_81}
{Toomre} A., 1981, in Structure and Evolution of Normal Galaxies, {Fall} S.~M.,
  {Lynden-Bell} D., eds., pp. 111--136

\bibitem[{{Vanderplas} {et~al}\mbox{.}(2012){Vanderplas}, {Connolly},
  {Ivezi{\'c}}, \& {Gray}}]{astroML}
{Vanderplas} J., {Connolly} A., {Ivezi{\'c}} {\v Z}., {Gray} A., 2012, in
  Conference on Intelligent Data Understanding (CIDU), pp. 47 --54

\bibitem[{{Willett} {et~al}\mbox{.}(2013){Willett}, {Lintott}, {Bamford},
  {Masters}, {Simmons}, {Casteels}, {Edmondson}, {Fortson}, {Kaviraj}, {Keel},
  {Melvin}, {Nichol}, {Raddick}, {Schawinski}, {Simpson}, {Skibba}, {Smith}, \&
  {Thomas}}]{Willett_13}
{Willett} K.~W. {et~al.}, 2013, MNRAS, 435, 2835

\bibitem[{{Willett} {et~al}\mbox{.}(2015){Willett}, {Schawinski}, {Simmons},
  {Masters}, {Skibba}, {Kaviraj}, {Melvin}, {Wong}, {Nichol}, {Cheung},
  {Lintott}, \& {Fortson}}]{Willett_15}
{Willett} K.~W. {et~al.}, 2015, MNRAS, 449, 820

\end{thebibliography}

\end{document}